\documentclass[onecolumn]{IEEEtran}
\ifCLASSINFOpdf
\else
\fi

\usepackage{graphicx}
\usepackage{amsmath}
\usepackage{amssymb}
\usepackage{graphics}
\usepackage{latexsym}
\usepackage[dvips]{epsfig}
\usepackage[nospace]{cite}
\usepackage{subfigure}
\usepackage{tabularx}
\usepackage{booktabs}
\usepackage{color}

\newtheorem{Definition}{Definition}
\newtheorem{Lemma}{Lemma}

\newtheorem{Proposition}[Lemma]{Proposition}
\newtheorem{Theorem}{Theorem}

\newtheorem{Remark}{Remark}

\def\Pr{{\rm {Pr}}}
\def\E{{\rm  E}}
\def\Var{{\rm {Var}}}

\allowdisplaybreaks[1]

\begin{document}
%
\title{Achievable Rates for Gaussian Degraded Relay Channels with Non-Vanishing Error Probabilities}
%
%
%


\author{Silas~L.~Fong and Vincent~Y.~F.~Tan
\thanks{S.~L.~Fong and V.~Y.~F.~Tan are supported in part by the NUS Young Investigator award (grant number R-263-000-B37-133) and a Ministry of Education Tier 2 grant (grant number R-263-000-B61-113).}
\thanks{Silas~L.~Fong and Vincent~Y.~F.~Tan are with the Department of Electrical and Computer Engineering, National University of Singapore, Singapore (e-mail: \texttt{\{silas\_fong,vtan\}@nus.edu.sg}).}}
\maketitle

\begin{abstract}
This paper revisits the Gaussian degraded relay channel, where the link that carries information from the source to the destination is a physically degraded version of the link that carries information from the source to the relay. The source and the relay are subject to expected power constraints. The $\varepsilon$-capacity of the channel is characterized and it is strictly larger than the capacity for any $\varepsilon>0$, which implies that the channel does not possess the strong converse property. The proof of the achievability part is based on several key ideas: block Markov coding which is used in the classical decode-forward strategy, power control for Gaussian channels under expected power constraints, and a careful scaling between the block size and the total number of block uses. The converse part is proved by first establishing two non-asymptotic lower bounds on the error probability, which are derived from the type-II errors of some binary hypothesis tests. Subsequently, each lower bound is simplified by conditioning on an event related to the power of some linear combination of the codewords transmitted by the source and the relay. Lower and upper bounds on the second-order term of the optimal coding rate are also obtained.
\end{abstract}


%
\IEEEpeerreviewmaketitle

\section{Introduction}\label{introduction}
\begin{figure}[t]
\centering
\includegraphics[width=2 in, height=0.7 in, bb = 203 321 563 451, angle=0, clip=true]{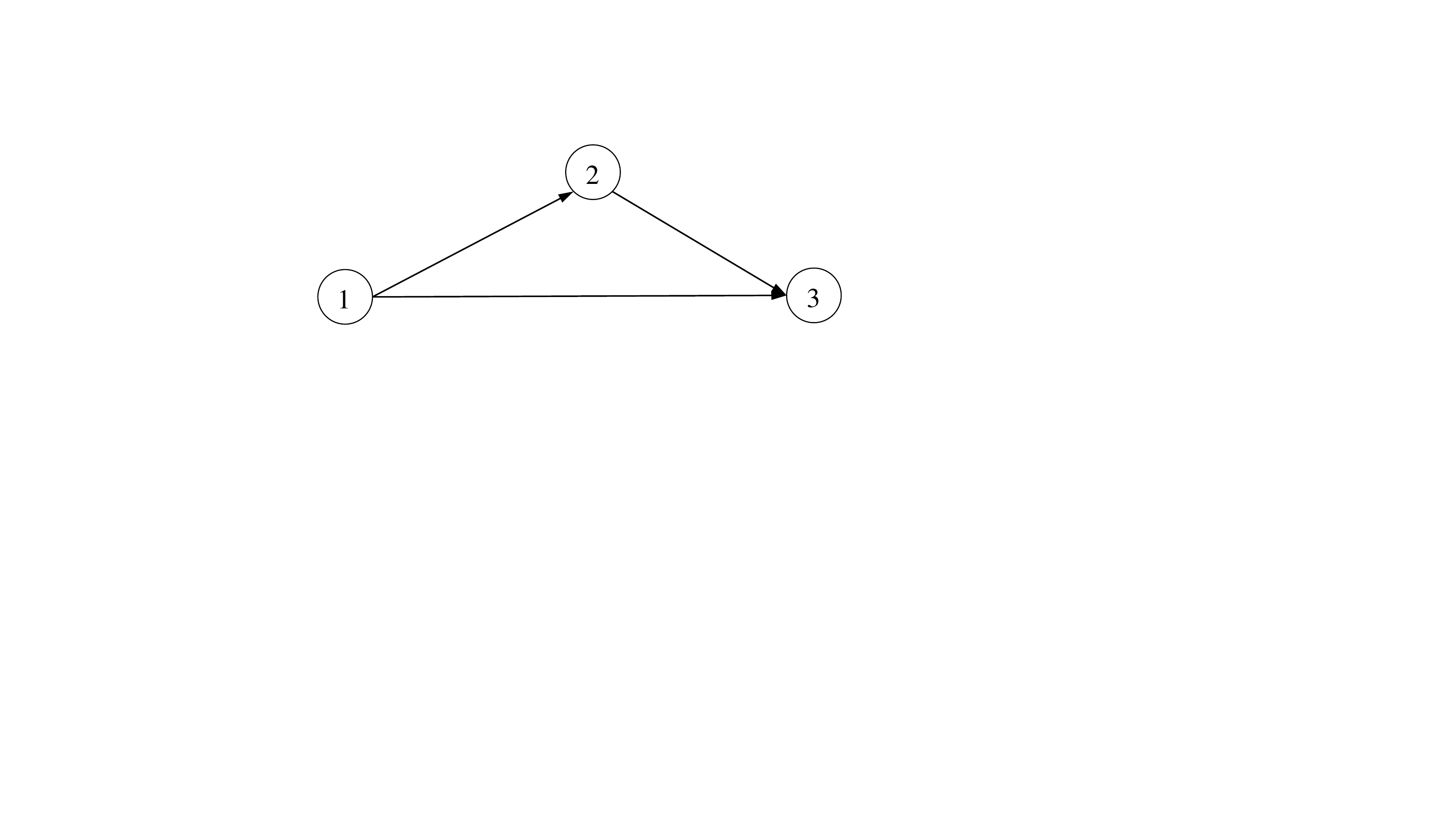}
\caption{A relay channel.}
\label{figureRC}
\end{figure}

%
%
%
%
\IEEEPARstart{T}{his} paper considers a relay channel (RC)~\cite{CEG} as illustrated in Figure~\ref{figureRC}, where nodes~$1$, $2$ and~$3$ denote the source, relay and destination respectively. Node~1 wants to transmit information to node~3 through node~2. The link that carries information from node~1 to node~3 is assumed to be a physically degraded version of the link that carries information from node~1 to node~2, and the RC described above is known as the \textit{degraded RC} in the literature \cite{CoverBook,elgamal}.
For the discrete memoryless degraded RC where the alphabets of the input variables $X_1$ and $X_2$ and the output variables $Y_2$ and $Y_3$ are finite, the channel characterized by a transition matrix $q_{Y_2,Y_3|X_1, X_2}$ satisfies
\begin{equation*}
q_{Y_2,Y_3|X_1, X_2} = q_{Y_2|X_1, X_2}q_{Y_3|X_2, Y_2}.
\end{equation*}
The capacity of the discrete memoryless degraded RC was shown in~\cite[Th.~1]{CEG} to be
\begin{equation}
\max_{p_{X_1, X_2}}\min\{I(X_1; Y_2|X_2), I(X_1, X_2; Y_3)\}. \label{DMcapacity}
\end{equation}
For the Gaussian degraded RC which is the main focus of this paper, the constituent channels $q_{Y_2|X_1, X_2}$ and $q_{Y_3|X_2, Y_2}$ are given by
\begin{equation*}
Y_{2} = X_{1} + Z_{2}
\end{equation*}
 and
 \begin{align*}
 Y_{3} &= X_{2} + Y_{2} + Z_{3}
 \end{align*}
  respectively, where $Z_2$ and $Z_3$ are independent zero-mean Gaussian random variables whose variances are denoted by $N_2>0$ and $N_3>0$ respectively.
If we let $P_1>0$ and $P_2>0$ denote the admissible power used by nodes~$1$ and~$2$ respectively, then
the capacity was shown in~\cite[Th.~5]{CEG} to be
\begin{equation}
C(P_1, P_2)\triangleq \max_{0\le \alpha\le 1}\min\left\{\mathrm{C}\left(\frac{\alpha P_1}{N_2}\right),\mathrm{C}\left(\frac{P_1 + P_2 +2\sqrt{(1-\alpha)P_1 P_2}}{N_2+N_3}\right) \right\} \label{introEq1}
\end{equation}
where
\begin{equation}
\mathrm{C}(x)\triangleq \frac{1}{2}\log(1+x) \label{defCapacityFunction}
 \end{equation}
 denotes the capacity of the additive white Gaussian noise (AWGN) channel with signal-to-noise ratio $x>0$. The capacities in~\eqref{DMcapacity} and~\eqref{introEq1} coincide with the cut-set outer bounds for the discrete memoryless model~\cite[Sec.~15.7]{CoverBook} and the Gaussian model~\cite[Sec.~15.1.4]{CoverBook} respectively.

  Although the capacity of the degraded RC is well known, it only characterizes the maximum achievable rate with \emph{vanishing} error probability. The maximum achievable rate with {\em non-vanishing} error probability for the degraded (discrete or Gaussian) RC has not been investigated previously. Recall that the $\varepsilon$-capacity~\cite{PPV10} is the maximum achievable rate with asymptotic average error probability no larger than $\varepsilon$. Due to the importance of communications in the presence of relays in large networks, we are motivated to revisit the fundamental limits of communicating over the Gaussian RC. As the capacity of the Gaussian RC is still unknown, we study a simpler model, the Gaussian degraded RC. The study of the $\varepsilon$-capacity and second-order asymptotics \cite{PPV10, TanBook} are of fundamental importance in today's latency- and delay-limited communication systems \cite{PPV10}. This is particularly true for systems where a tradeoff between rate and error probability is possible (due to the absence of the strong converse). Therefore, we investigate the first-order tradeoff for the Gaussian degraded RC by studying the $\varepsilon$-capacity in this paper. As we will see in the next subsection, our $\varepsilon$-capacity result implies that code designers can indeed operate at rates above the capacity and arbitrarily close to the $\varepsilon$-capacity if they can tolerate a non-zero error probability~$\varepsilon$. Furthermore, the bounds on the second-order asymptotics provide approximations to the non-asymptotic fundamental limits of the Gaussian degraded RC. This is the first work that studies relay channels non-asymptotically.
\subsection{Main Contributions}\label{secMainContribution}
In this paper, we investigate the Gaussian degraded RC under expected power constraints at both the source and the relay and fully characterize the $\varepsilon$-capacity to be
\begin{equation}
C_\varepsilon = C\left(\frac{P_1}{1-\varepsilon}, \frac{P_2}{1-\varepsilon}\right). \label{introEq2}
\end{equation}
Comparing~\eqref{introEq1} with~\eqref{introEq2}, we see that the $\varepsilon$-capacity is strictly increasing in~$\varepsilon$ and strictly larger than the capacity for any $\varepsilon\in (0,1)$, which implies that the Gaussian degraded RC does not admit the strong converse property possessed by the discrete memoryless channel (DMC) \cite[Theorem~2]{Wolfowitz57}, the AWGN channel~\cite{Yoshihara}, and many other classes of memoryless channels~\cite{Han10}. The proof of the achievability part is based on the ideas of power control for Gaussian channels under expected power constraints~\cite{YCDP15,TFT15}, the decode-forward strategy for the RC~\cite{CEG,elgamal}, multiple applications of the Shannon's threshold decoding bound~\cite{sha57},\cite[Th.~2]{PPV10}, and the non-asymptotic packing lemma~\cite{verdu12}. The converse part is proved by first establishing two non-asymptotic lower bounds on the error probability, which are derived from the type-II errors of appropriately-defined binary hypothesis tests. Each lower bound is then simplified by conditioning on an event related to the power of some linear combination of the codewords transmitted by the source and the relay.

In addition, we obtain lower and upper bounds on the second-order term of the optimal coding rate, which is formally defined as follows: For any $\varepsilon\in(0,1)$ and any $n\in \mathbb{N}$, let $M^*(n, \varepsilon, P_1, P_2)$ be maximum size of the message set that can be supported by a length-$n$ code whose average probability of error is no larger than $\varepsilon$ and whose  admissible powers are $P_1$ and $P_2$. Then,   the \emph{second-order term of the asymptotic expansion of $\log M^*(n, \varepsilon, P_1, P_2)$} is
\begin{equation*}
\theta_{n, \varepsilon}\triangleq \log M^*(n, \varepsilon, P_1, P_2) - nC_\varepsilon.
 \end{equation*}
 In other words, $\theta_{n, \varepsilon}$ denotes $n$ times the minimum backoff from the  $\varepsilon$-capacity over all length-$n$ codes. A by-product of our proof techniques yields
 \begin{align*}
 \liminf_{n\rightarrow \infty}\frac{\theta_{n, \varepsilon}}{n^{4/5}}\ge -\underline{\Psi}(P_1, P_2, N_1, N_2, \varepsilon)
 \end{align*}
and
 \begin{align*}
 \limsup_{n\rightarrow \infty}\frac{\theta_{n, \varepsilon}}{\sqrt{n}\log n}\le \overline{\Psi}(P_1, P_2, N_1, N_2, \varepsilon)
 \end{align*}
 for some positive constants $\underline{\Psi}(P_1, P_2, N_1, N_2, \varepsilon)$ and $\overline{\Psi}(P_1, P_2, N_1, N_2, \varepsilon)$. While the exact scaling of~$\theta_{n, \varepsilon}$ is still unknown at this point, we have attempted to optimize them by, for example, carefully balancing the number of blocks used for the decode-forward strategy and the number of channel uses per block.

\subsection{Related Work}
The capacity of the Gaussian degraded RC was first proved by Cover and El~Gamal in their seminal paper on RCs~\cite{CEG}. This paper is concerned with refined asymptotics of achievable rates for this channel. Generally, there are two main asymptotic regimes of interest when one seeks to obtain refined estimates of achievable rates or achievable error probabilities for communication: (i) The error exponent regime where the rate is fixed below capacity and one is interested in the exponential rate of decay of the error probability; (ii) The non-vanishing error regime where one is also possibly concerned with the second-order asymptotics in addition to the $\varepsilon$-capacity. For the former, Bradford and Laneman~\cite{Bra12} and Tan~\cite{Tan15} derived bounds on the error exponent (reliability function) of the discrete memoryless RC. Also see~\cite{LG06, ZM07, YKG10} for other related works on error exponents for RCs. For the latter, there is a body of work for other multi-terminal, one-hop channel models~\cite{TanBook} but this is the first work that systematically studies the non-vanishing error probability asymptotics for a specific multi-hop channel model. Our converse technique is closely related to that used to establish converses for single- and multi-user Gaussian channels with feedback~\cite{TFT15}. Similarly, the upper bound on the error exponent obtained by Tan~\cite{Tan15} for the discrete memoryless RC is closely related to Haroutunian's exponent for DMCs with feedback~\cite{Haroutunian77}.


\subsection{Paper Outline}
This paper is organized as follows. The notations used in this paper are described in the next subsection. Section~\ref{sectionDefinition} presents the problem formulation of the Gaussian degraded RC and its~$\varepsilon$-capacity, which is the main result in this paper. The preliminaries for the proof of the main result are contained in Section~\ref{sectionPrelim}, which includes a non-asymptotic packing lemma and two non-asymptotic bounds derived from the type-II errors of binary hypothesis tests. Sections~\ref{secAchievability} and~\ref{secConverse} present the achievability and converse parts respectively of the proof of the main result.
\subsection{Notation}\label{notation}
We will take all logarithms to base $e$, and we will use the convention that $0\log 0=0$ and $0\log \frac{0}{0}=0$ throughout this paper. For any mapping $g:\mathcal{X}\rightarrow\mathcal{Y}$ and any $\mathcal{S}\subseteq \mathcal{Y}$, we define $g^{-1}(\mathcal{S})\triangleq\{x\in \mathcal{X}|\, g(x)\in \mathcal{S}\}$. The set of natural, real and non-negative real numbers are denoted by $\mathbb{N}$, $\mathbb{R}$ and $\mathbb{R}_+$ respectively. The $n$-dimensional all-zero and all-one tuples are denoted by $0^n$ and $1^n$ respectively. The Euclidean norm of a tuple $x^n\in \mathbb{R}^n$ is denoted by $\|x^n\|\triangleq \sqrt{\sum_{k=1}^n x_k^2}$.

We use $\Pr\{\mathcal{E}\}$ to represent the probability of an
event~$\mathcal{E}$, and we let $\mathbf{1}\{\mathcal{E}\}$ be the characteristic function of $\mathcal{E}$. We use an upper case letter (e.g.,~$X$) to denote a random variable (with alphabet $\mathcal{X}$), and use the corresponding lower case letter (e.g., $x$) to denote a realization of the random variable.
We use $X^n$ to denote a random tuple $(X_1,  X_2,  \ldots,  X_n) \in\mathcal{X}^n$. 
We let $p_X$ and $p_{Y|X}$ denote the probability distribution of $X$ and the conditional probability distribution of $Y$ given $X$ respectively for any random variables~$X$ and~$Y$.
We let $p_Xp_{Y|X}$ denote the joint distribution of $(X,Y)$, i.e., $p_Xp_{Y|X}(x,y)=p_X(x)p_{Y|X}(y|x)$ for all $x$ and $y$.
To make the dependence on the distribution explicit, we often let $\Pr_{p_X}\{ g(X)\in\mathcal{A}\}$ denote $\int_{\mathcal{X}} p_X(x)\mathbf{1}\{g(x)\in\mathcal{A}\}\, \mathrm{d}x$ for any set $\mathcal{A}\subseteq \mathbb{R}$ and any real-valued $g$ whose domain includes~$\mathcal{X}$.
The expectation and the variance of~$g(X)$ are denoted as
$
\E_{p_X}[g(X)]$ and
$
 \Var_{p_X}[g(X)]=\E_{p_X}[(g(X)-\E_{p_X}[g(X)])^2]$
 respectively. For simplicity, we drop the subscript of a notation if there is no ambiguity. 
We let $\mathcal{N}(\,\cdot\,;\mu,N): \mathbb{R}^n \rightarrow [0,\infty)$ be the joint probability density function of~$n$ independent copies of the standard Gaussian random variable, i.e.,
   \begin{equation*}
  \mathcal{N}(z^n;\mu,N) = \frac{1}{(2\pi N)^{\frac{n}{2}}}e^{-\sum\limits_{k=1}^n\frac{ (z_k-\mu)^2}{2N}}.
  \end{equation*}


\section{Gaussian degraded Relay Channel and Its~$\varepsilon$-Capacity} \label{sectionDefinition}
\begin{figure}[t]
\centering
\includegraphics[width=2.8 in, height=1.1 in, bb = 200 281 729 497, angle=0, clip=true]{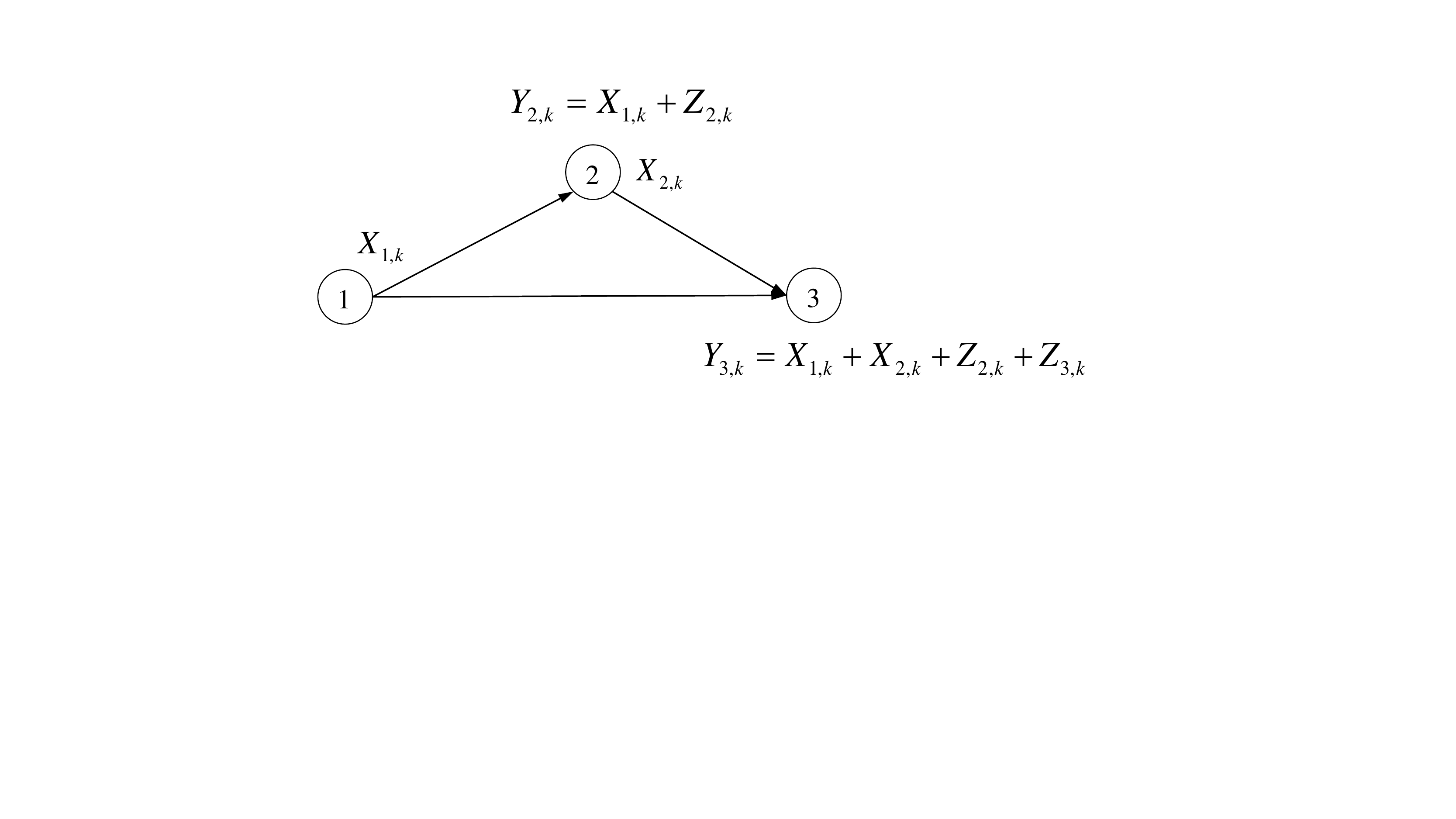}
\caption{Gaussian degraded RC.}
\label{figureGaussianRC}
\end{figure}
We consider the Gaussian degraded RC as illustrated in Figure~\ref{figureGaussianRC}, where nodes~$1$, $2$ and~$3$ denote the source, relay and destination respectively.
Node~1 transmits information to node~3 in~$n$ channel uses as follows.
Node~1 chooses a message $W$
destined for node~3. For the $k^{\text{th}}$ channel use for each $k\in \{1, 2, \ldots, n\}$, node~1 and node~2 transmit $X_{1,k} \in \mathbb{R}$ and $X_{2,k} \in \mathbb{R}$ respectively while node~2 and node~3 receive
\begin{equation}
Y_{2,k} = X_{1,k} + Z_{2,k} \label{introY2k}
\end{equation}
 and
 \begin{align}
 Y_{3,k} &= X_{2,k} + Y_{2,k} + Z_{3,k}  \label{introY3k} \\
& \stackrel{\eqref{introY2k}}{=} X_{1,k}+X_{2,k}+Z_{2,k}+Z_{3,k} \notag
 \end{align}
  respectively\footnote{Throughout this paper, the equation number above a binary operation explains why the binary operation holds.},
 where $Z_2^n\sim \mathcal{N}(z_2^n; 0, N_2)$ and $Z_3^n\sim \mathcal{N}(z_3^n; 0, N_3)$ are independent Gaussian random tuples which denote the noises received at node~2 and node~3 respectively. In addition, $X_1^n$ is a function of~$W$ and $X_{2,k}$ is a function of $Y_2^{k-1}$ for each $k\in\{1, 2, \ldots, n\}$. Node~1 and node~2 are subject to the following expected power constraints for some fixed $P_1>0$ and $P_2>0$:
 \begin{equation}
 \E\left[\frac{1}{n}\sum_{k=1}^n X_{i,k}^2\right]\le P_i \label{powerConstraint}
 \end{equation}
for each $i\in\{1,2\}$. After~$n$ channel uses, node~3 declares~$\hat W$ to be the
transmitted~$W$ based on $Y_3^n$. The RC described above is known as the \emph{Gaussian degraded RC} \cite[Sec.~4]{CEG} (see also \cite[Sec.~15.1.4]{CoverBook}).

The following five standard definitions formally define a Gaussian degraded RC and its $\varepsilon$-capacity.
\medskip
\begin{Definition} \label{defCode}
An $(n, M, P_1, P_2)$-code consists of the following:
\begin{enumerate}
\item A message set
\begin{equation*}
\mathcal{W}=\{1, 2, \ldots, M\}
\end{equation*}
 at node~1. Message $W$ is uniform on $\mathcal{W}$.

\item An encoding function
\begin{equation*}
f_{1,k}: \mathcal{W} \rightarrow \mathbb{R}
 \end{equation*}
at node~1 for each $k\in\{1, 2, \ldots, n\}$ such that
\begin{equation}
X_{1,k}=f_{1,k}(W). \label{sourceEncoder}
\end{equation}
In addition, the power constraint~\eqref{powerConstraint} must be satisfied for~$i=1$.

\item An encoding function
\begin{equation*}
f_{2,k}: \mathbb{R}^{k-1} \rightarrow \mathbb{R}
 \end{equation*}
at node~2 for each $k\in\{1, 2, \ldots, n\}$ such that
\begin{equation}
X_{2,k}=f_{2,k}(Y_2^{k-1}). \label{relayEncoder}
\end{equation}
In addition, the power constraint~\eqref{powerConstraint} must be satisfied for~$i=2$.

\item A decoding function
\begin{equation*}
\varphi : \mathbb{R}^n \rightarrow \mathcal{W}
 \end{equation*}
at node~3 such that
 \begin{equation*}
 \hat W = \varphi(Y_3^n).
 \end{equation*}
\end{enumerate}
\end{Definition}
\medskip
\begin{Definition}\label{defchannel}
A Gaussian degraded RC is characterized by the probability density function $q_{Y_2, Y_3|X_1, X_2}$ satisfying
\begin{align}
q_{Y_2, Y_3|X_1, X_2}(y_2, y_3|x_1, x_2)& =q_{Y_2|X_1}(y_2|x_1)q_{Y_3|X_2, Y_2}(y_3|x_2, y_2) \notag\\
 &=\mathcal{N}(y_2-x_1; 0, N_2)\mathcal{N}(y_3-y_2-x_2; 0, N_3) \label{defChannelInDefinition}
\end{align}
for some $N_2>0$ and $N_3>0$ such that the following holds for any $(n, M, P_1, P_2)$-code: For each $k\in\{1, 2, \ldots, n\}$,
\begin{align}
p_{W, X_1^k, X_2^k, Y_2^k, Y_3^k}
 = p_{W, X_1^k, X_2^k, Y_2^{k-1}, Y_3^{k-1}}p_{Y_{2,k}, Y_{3,k}|X_{1,k}, X_{2,k}}
 \label{memorylessStatement*}
\end{align}
where
\begin{equation}
p_{Y_{2,k}, Y_{3,k}|X_{1,k}, X_{2,k}}(y_{2,k}, y_{3,k}|x_{1,k}, x_{2,k}) = q_{Y_{2}, Y_{3}|X_{1}, X_{2}}(y_{2,k}, y_{3,k}|x_{1,k}, x_{2,k}) \label{defChannelInDefinition*}
\end{equation}
for all $x_{1,k}$, $x_{2,k}$, $y_{2,k}$ and $y_{3,k}$.
Since $p_{Y_{2,k}, Y_{3,k}|X_{1,k}, X_{2,k}}$ does not depend on~$k$ by \eqref{defChannelInDefinition*}, the channel is stationary.
\end{Definition}
\medskip

For any $(n, M, P_1, P_2)$-code defined on the Gaussian degraded RC, let $p_{W, X_1^n, X_2^n, Y_2^n, Y_3^n, \hat W}$ be the joint distribution induced by the code. Then, we can use Definitions~\ref{defCode} and~\ref{defchannel} to express $p_{W, X_1^n, X_2^n, Y_2^n, Y_3^n, \hat W}$ as follows:
\begin{align}
&p_{W, X_1^n, X_2^n, Y_2^n, Y_3^n, \hat W} \notag\\
&\stackrel{\text{(a)}}{=} p_{W, X_1^n, X_2^n, Y_2^n, Y_3^n}p_{\hat W |Y_3^n} \label{memorylessStatement} \\
& \stackrel{\text{(c)}}{=} p_W \left(\prod_{k=1}^n  p_{X_{1,k}|W,Y_2^{k-1}, Y_3^{k-1}} p_{X_{2,k}|Y_2^{k-1}, Y_3^{k-1}} p_{Y_{2,k}|X_{1,k}} p_{Y_{3,k}|X_{2,k}, Y_{2,k}}\right)p_{\hat W |Y_3^n}.\label{memorylessStatement**}
\end{align}
\medskip
\begin{Definition} \label{defError}
For an $(n, M, P_1, P_2)$-code defined on the Gaussian degraded RC, we can calculate according to~\eqref{memorylessStatement**} the average probability of decoding error
$
\Pr\{W\ne \hat W\}$.
We call an $(n, M, P_1, P_2)$-code with average probability of decoding error no larger than $\varepsilon$ an $(n, M, P_1, P_2, \varepsilon)$-code.
\end{Definition}
\medskip
\begin{Definition} \label{defAchievableRate}
Fix an $\varepsilon\in(0,1)$. A rate $R\ge 0$ is \textit{$\varepsilon$-achievable} for the Gaussian degraded RC if there exists a sequence of $(n, M_n, P_1, P_2, \varepsilon)$-codes such that
\begin{equation*}
\liminf\limits_{n\rightarrow \infty}\frac{1}{n}\log M_n \ge R.
\end{equation*}
\end{Definition}
\medskip
\begin{Definition}\label{defCapacity}
For each $\varepsilon\in(0,1)$, the \textit{$\varepsilon$-capacity} of the Gaussian degraded RC is defined as
\begin{equation*}
C_\varepsilon \triangleq \sup\left\{R\left|R\text{ is $\varepsilon$-achievable}\right.\right\}.
\end{equation*}
 The \textit{capacity} is defined as
 \begin{equation*}
  C_0 \triangleq \inf_{\varepsilon>0} C_\varepsilon = \lim_{\varepsilon\rightarrow 0} C_\varepsilon.
  \end{equation*}
\end{Definition}
\medskip

Recall the definition of $\mathrm{C}(\cdot)$ in~\eqref{defCapacityFunction} and define
   \begin{align}
 R_{\text{cut-set}}\left(\alpha, P_1, P_2\right)\triangleq \min\left\{\mathrm{C}\left(\frac{\alpha P_1}{N_2}\right),\mathrm{C}\left(\frac{P_1 + P_2 +2\sqrt{(1-\alpha)P_1 P_2}}{N_2+N_3}\right) \right\}. \label{defRcutset}
 \end{align}
  It is well known \cite[Sec.~15.1.4]{CoverBook} that the capacity of the Gaussian degraded RC coincides with the cut-set bound, i.e.,
\begin{equation}
C_0 = \max\limits_{0\le\alpha\le 1}  R_{\text{cut-set}}(\alpha, P_1, P_2). \label{eqnCapacityC}
\end{equation}
The following theorem is the main result in this paper. The proof of the main result consists of an achievability part and a converse part, which will be presented in Section~\ref{secAchievability} and Section~\ref{secConverse} respectively.
\medskip
\begin{Theorem} \label{thmMainResult}
Fix an $\varepsilon \in (0,1)$. Then,
\begin{equation}
C_\varepsilon =  \max\limits_{0\le\alpha\le 1}R_{\text{cut-set}}\left(\alpha, \frac{P_1}{1-\varepsilon}, \frac{P_2}{1-\varepsilon}\right). \label{thmMainResultSt}
\end{equation}
\end{Theorem}
\begin{Remark}\label{remark1}
Theorem~\ref{thmMainResult} fully characterizes the $\varepsilon$-capacity of the Gaussian degraded RC, which depends on~$\varepsilon$ and is stricter larger than the capacity in view of~\eqref{eqnCapacityC} and~\eqref{thmMainResultSt}. In other words, the Gaussian degraded RC subject to expected power constraints at the source and the relay does not possess the strong converse property.
\end{Remark}
\medskip
\begin{Remark}
\begin{figure}[t]
\centering
\includegraphics[width=2.7 in, height=1.1 in, bb = 202 270 782 500, angle=0, clip=true]{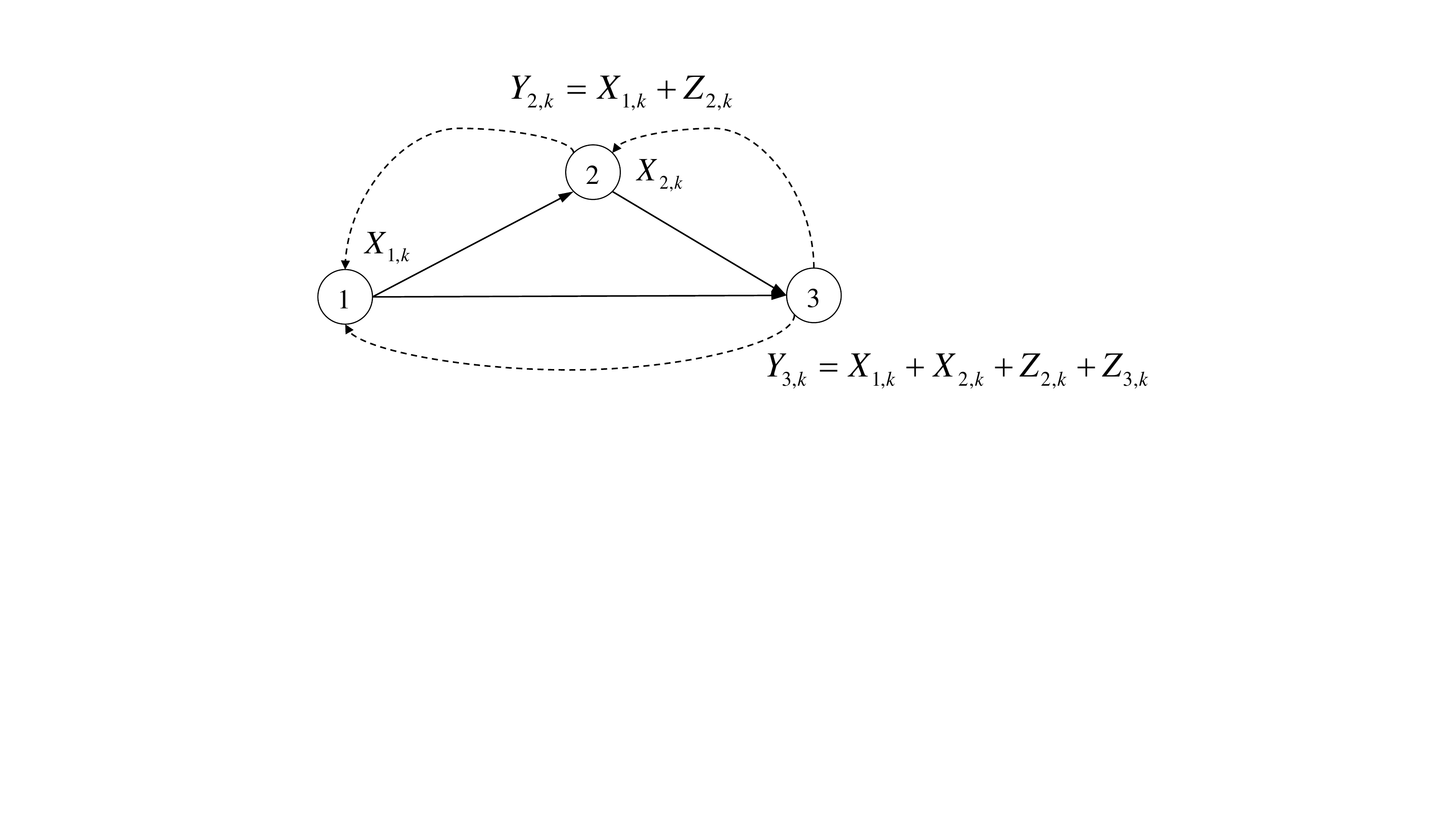}
\caption{Gaussian degraded RC with feedback links indicated by dashed lines.}
\label{figureGaussianRCFB}
\end{figure}
Define an $(n, M, P_1, P_2, \varepsilon)$-feedback code in a similar way as done in Definitions~\ref{defCode} and~\ref{defError} except that $(Y_2^{k-1}, Y_3^{k-1})$ are also available for encoding $X_{1,k}$ and $X_{2,k}$, i.e., $X_{1,k}=f_{1,k}(W, Y_2^{k-1}, Y_3^{k-1})$ and $X_{2,k}=f_{2,k}(Y_2^{k-1}, Y_3^{k-1})$ respectively (compare to~\eqref{sourceEncoder} and~\eqref{relayEncoder}). In other words, the encoding operations of the $(n, M, P_1, P_2, \varepsilon)$-feedback code assume the presence of three perfect feedback links that carry the outputs $(Y_2^{k-1}, Y_3^{k-1})$ to the source and the relay as illustrated in Figure~\ref{figureGaussianRCFB}. Similar to Definition~\ref{defCapacity}, we define the feedback $\varepsilon$-capacity $C_{\varepsilon}^{\text{FB}}$ to be the supremum of rates achievable by all sequences of $(n, M, P_1, P_2, \varepsilon)$-feedback codes. Then the same converse proof of Theorem~\ref{thmMainResult} presented in Section~\ref{secConverse} can be used to show that
\begin{equation*}
C_{\varepsilon}^{\text{FB}}\le \max\limits_{0\le\alpha\le 1}R_{\text{cut-set}}\left(\alpha, \frac{P_1}{1-\varepsilon}, \frac{P_2}{1-\varepsilon}\right)
 \end{equation*}
 for all $\varepsilon\in(0,1)$, which implies from Theorem~\ref{thmMainResult} that the presence of the perfect feedback links does not increase the $\varepsilon$-capacity. It is well known the presence of the perfect feedback links does not increase the capacity~\cite[Sec.~V]{CEG}. Our work shows that this phenomenon also holds for the $\varepsilon$-capacity for all $\varepsilon\in(0,1)$.
\end{Remark}
\medskip
\begin{Remark}\label{remark2}
In addition to using the block Markov coding technique in the classical decode-forward strategy~\cite{CEG,elgamal}, another key ingredient in the achievability proof of Theorem~\ref{thmMainResult} presented in Section~\ref{secAchievability} is the careful control of the expected power of transmitted codewords by means of allocating zero power to a deterministic subset of the message set (cf.\ Section~\ref{subsec3}). This simple power allocation idea is known as \emph{power control}\cite{YCDP15}. We also carefully scale the number of blocks used for the decode-forward strategy and number of channel uses per block to optimize the backoff from the capacity.
\end{Remark}
\medskip
\begin{Remark}\label{remark1*}
For each $\varepsilon\in(0,1)$ and each\footnote{Since~$n$ is used to denote the number of channel uses in each block under the block Markov coding strategy in the achievability proof of Theorem~\ref{thmMainResult}, we use~$m$ instead of~$n$ to denote the total number of channel uses in this remark in order to avoid confusion.} $m\in\mathbb{N}$, let
\begin{equation}
M^*(m, \varepsilon, P_1, P_2) \triangleq \max\left\{M\in\mathbb{N}\left|\text{There exists an $(m, M, P_1, P_2, \varepsilon)$-code}\right.\right\} \label{defM*inRemark}
\end{equation}
 be the maximum size of the message alphabet that can be supported by a length-$m$ code whose average probability of error is no larger than~$\varepsilon$.
Then, the bound~\eqref{achRateEq8} in the achievability proof of Theorem~\ref{thmMainResult} implies the following: There exists a positive number $c$ which is a function of $(\varepsilon, P_1, P_2, N_1, N_2)$ and does not depend on~$m$ such that for all $m\in\mathbb{N}$,
\begin{align}
\log M^*(m, \varepsilon, P_1, P_2) \ge m C_\varepsilon - c\,m^{4/5}\,. \label{nonAsymptoticBound}
\end{align}
The backoff term~$- c\,m^{4/5}$ is due to the interplay of the following three factors in our proposed decode-forward strategy:
\begin{enumerate}
\item[(i)] Each message is divided into~$(m^{1/5}-1)$ submessages. With probability $p_{\text{erased}}\triangleq\varepsilon + O(m^{-1/5})$, the message is erased such that the source transmits almost nothing in the entire~$m$ channel uses. With probability $1-p_{\text{erased}}$, every submessage is transmitted through a length-$m^{4/5}$ block code using the decode-forward strategy. By the decode-forward strategy, the $(m^{1/5}-1)$ submessages are transmitted to the destination in~$m$ channel uses where the last length-$m^{4/5}$ block contains no new information, which contributes to part of the backoff term~$- c\,m^{4/5}$.
         \item[(ii)] Each non-erased submessage is encoded using i.i.d.\ Gaussian codewords with variances slightly backed off from our designed admissible peak power~$\frac{P_i}{1-\varepsilon + O(m^{-1/5})}$ by another factor of $(1-m^{-1/5})$ so that the probability of violating the peak power~$\frac{P_i}{1-\varepsilon + O(m^{-1/5})}$ associated with each non-erased submessage is less than\footnote{It can be verified by applying Chebyshev's inequality to each length-$m^{4/5}$ block.} $O(m^{-2/5})$, which ensures that the probability of violating the peak power~$\frac{P_i}{1-\varepsilon + O(m^{-1/5})}$ for each non-erased message is less than $O(m^{1/5})\times O(m^{-2/5}) = O(m^{-1/5})$. The backoff factor $(1-m^{-1/5})$ contributes to part of the backoff term~$- c\,m^{4/5}$.
              \item[(iii)] The decoding error probability for each non-erased submessage scales as $O(m^{-2/5})$, which ensures that the decoding error probability for each non-erased message is less than $O(m^{1/5})\times O(m^{-2/5}) = O(m^{-1/5})$.
            \end{enumerate}
In view of (i) and (ii), we can see that the expected power consumed at the source and the relay are approximately~$P_1$ and~$P_2$ respectively. In view of (i) and (iii), we can see that the error probability is approximately~$\varepsilon$. Finally, the backoff term~$- c\,m^{4/5}$ in \eqref{nonAsymptoticBound} is a consequence of~(i) and~(ii).
\end{Remark}

\medskip
\begin{Remark}\label{remark3}
The converse proof of Theorem~\ref{thmMainResult} presented in Section~\ref{secConverse} consists of two steps. First, we establish two non-asymptotic lower bounds on the error probability which are derived from the type-II errors of some binary hypothesis tests. Second, we simplify each lower bound by conditioning on an event related to the power of some linear combination of $X_1^n$ and $X_2^n$. These events are formally defined in~\eqref{defEventEIConverse}. The final bound of the converse proof in~\eqref{converseProofEq22} implies the following for each $\varepsilon\in(0,1)$: There exists a positive number $c$ which is a function of $(\varepsilon, P_1, P_2, N_1, N_2)$ (but does not depend on~$n$) such that for all $n\in\mathbb{N}$,
\begin{align}
\log M^*(n, \varepsilon, P_1, P_2) \le n C_\varepsilon + c\,\sqrt{n}\log n \label{nonAsymptoticBound*}
\end{align}
where $M^*(n, \varepsilon, P_1, P_2)$ is as defined in~\eqref{defM*inRemark}.
The upper bound on the second-order term implied from~\eqref{nonAsymptoticBound*} scales as $O(\sqrt{n}\log n)$ rather than the usual $O(\sqrt{n})$ for the point-to-point case that results from applying the central limit theorem to a non-asymptotic converse bound based on binary hypothesis testing. This is because for the Gaussian RC considered herein, we need to simplify the non-asymptotic bounds by additionally conditioning on the aforementioned events, which then results in a looser second-order term $O(\sqrt{n}\log n)$.
\end{Remark}
\medskip
\begin{Remark} \label{remark4}
Theorem~\ref{thmMainResult} completely characterizes $C_\varepsilon$ for all $\varepsilon \in (0,1)$ under the expected power constraints~\eqref{powerConstraint}. However, Theorem~\ref{thmMainResult} does not apply to the formulation where each node~$i\in\{1,2\}$ (source and relay) is subject to the peak power constraint
 \begin{equation}
 \Pr\left\{\frac{1}{n}\sum_{k=1}^n X_{i,k}^2\le P_i\right\}=1. \label{powerConstraintPeak}
 \end{equation}
The difficulty in extending Theorem~\ref{thmMainResult} to the peak power constraint formulation is due to the following two facts:
 \begin{enumerate}
 \item[(i)] The power control arguments used in the achievability proof of Theorem~\ref{thmMainResult} are no longer valid under the peak power constraint formulation.
      \item[(ii)] The proof techniques in the converse proof of Theorem~\ref{thmMainResult} do not yield a tighter bound if the expected power constraints~\eqref{powerConstraint} are replaced with the peak power constraints~\eqref{powerConstraintPeak}.
          \end{enumerate}
\end{Remark}
\medskip
\begin{Remark} \label{remark5}
Under the peak power constraint formulation~\eqref{powerConstraintPeak},
it is well known~\cite[Sec.~IV]{CEG} that the capacity is also $\max\limits_{0\le\alpha\le 1}  R_{\text{cut-set}}(\alpha, P_1, P_2)$. This paper does not attempt to characterize the $\varepsilon$-capacity under the peak power constraint formulation. We leave this as an open problem for future research.
\end{Remark}


\section{Preliminaries} \label{sectionPrelim}
Sections~\ref{subsecPackingLemma} and~\ref{subsecBHT} present preliminaries for the achievability and converse parts of the proof of Theorem~\ref{thmMainResult} respectively.
\subsection{Non-Asymptotic Packing Lemma}\label{subsecPackingLemma}
Similar to typical sets used in joint typicality decoding~\cite[Sec.~3.1.2]{elgamal}, we define for any given joint distribution $s_{X,Y,Z}$ the threshold decoding set
 \begin{align}
 \mathcal{T}_{s_{X,Y,Z}}^{(n)}(Y;Z|X)
\triangleq \left\{\parbox[c]{.85 in}{$(x^n, y^n, z^n)\in \vspace{0.04 in}\\ \mathbb{R}^n\times \mathbb{R}^n\times \mathbb{R}^n$}\,\left| \, \parbox[c]{4.75 in}{$\sum_{k=1}^n \log\left(\frac{s_{Y|X,Z}(y_k|x_k, z_k)}{s_{Y|X}(y_k|x_k)}\right) \ge  n \E\left[\log\left(\frac{s(Y|X,Z)}{s(Y|X)}\right)\right] - \sqrt{n^{\frac{3}{2}} \Var\left[\log\left(\frac{s(Y|X,Z)}{s(Y|X)}\right)\right]} $} \right.\right\} \label{defSetThreshold}
 \end{align}
 where we have adopted the shorthand notations
 \begin{align*}
 \E\left[\log\left(\frac{s(Y|X,Z)}{s(Y|X)}\right)\right] \triangleq \E_{s_{X,Y,Z}}\left[\log\left(\frac{s_{Y|X,Z}(Y|X,Z)}{s_{Y|X}(Y|X)}\right)\right]
 \end{align*}
 and
 \begin{align*}
  \Var\left[\log\left(\frac{s(Y|X,Z)}{s(Y|X)}\right)\right] \triangleq  \Var_{s_{X,Y,Z}}\left[\log\left(\frac{s_{Y|X,Z}(Y|X,Z)}{s_{Y|X}(Y|X)}\right)\right].
 \end{align*}
 Statement~\eqref{st1LemmaPacking} in the following lemma can be viewed as a non-asymptotic version of the packing lemma~\cite[Sec.~3.2]{elgamal}.
\medskip
 \begin{Lemma} \label{lemmaPacking}
Fix a $p_{X,Y,Z}$ and an~$M\in \mathbb{N}$. Define
 \begin{equation*}
 p_{X^n, Y^n, Z^n}(x^n, y^n,z^n)\triangleq \prod_{k=1}^n p_{X, Y, Z}(x_k, y_k, z_k)
  \end{equation*}
  for each $(x^n, y^n, z^n)\in \mathcal{X}^n\times \mathcal{Y}^n\times \mathcal{Z}^n$.
Let $\{(X^n(i), Y^n(i), Z^n(i))\}_{i=1}^M$ be $M$ i.i.d.\ random tuples such that \linebreak $(X^n(1), Y^n(1), Z^n(1))\sim p_{X^n, Y^n, Z^n}$, and let $p_{X^n(i), Y^n(i), Z^n(i)}$ denote the distribution of~$(X^n(i), Y^n(i), Z^n(i))$ for each $i\in\{1, 2, \ldots, M\}$. Then, we have
 \begin{align}
\Pr_{p_{X^n(1), Y^n(1), Z^n(1)}}\left\{(X^n(1), Y^n(1), Z^n(1))\notin \mathcal{T}_{p_{X,Y,Z}}^{(n)}(Y;Z|X)\right\} \le \frac{1}{n^{1/2}}. \label{st1LemmaPacking}
\end{align}
In addition,
\begin{align}
&\Pr_{\left(\prod\limits_{\ell=1}^M p_{X^n(\ell), Z^n(\ell)}\right)p_{Y^n(1)|X^n(1), Z^n(1)}}\left\{\bigcup_{j\in \{2, 3, \ldots, M\}}(X^n(1), Y^n(1), Z^n(j)) \in \mathcal{T}_{p_{X,Y,Z}}^{(n)}(Y;Z|X)\right\} \notag\\
& \le (M-1)e^{-n \E\left[\log\left(\frac{s(Y|X,Z)}{s(Y|X)}\right)\right] + \sqrt{n^{\frac{3}{2}} \Var\left[\log\left(\frac{s(Y|X,Z)}{s(Y|X)}\right)\right]}}\,. \label{stLemmaPacking}
\end{align}
 \end{Lemma}
 \begin{IEEEproof}
Using~\eqref{defSetThreshold} and Chebyshev's inequality, we have~\eqref{st1LemmaPacking}. In addition, by following almost identical steps in  the proof of the non-asymptotic packing lemma~\cite[Lemma~2]{Ver12}, we obtain \eqref{stLemmaPacking}.
 \end{IEEEproof}
%

\subsection{Binary Hypothesis Testing}\label{subsecBHT}
The following definition concerning the non-asymptotic fundamental limits of a simple binary hypothesis test is standard. See for example \cite[Section~III.E]{PPV10}.
\medskip
\begin{Definition}\label{defBHTDivergence}
Let $p_{X}$ and $q_{X}$ be two probability distributions on some common alphabet $\mathcal{X}$. Let
\[
\mathcal{Q}(\{0,1\}|\mathcal{X})\triangleq \{
r_{Z|X} \,|\, \text{$Z$ and $X$ assume values in $\{0,1\}$ and $\mathcal{X}$ respectively}\}
\]
be the set of randomized binary hypothesis tests between $p_{X}$ and $q_{X}$ where $\{Z=0\}$ indicates the test chooses $q_X$, and let $\delta\in [0,1]$ be a real number. The minimum type-II error in a simple binary hypothesis test between $p_{X}$ and $q_{X}$ with type-I error no larger than $1-\delta$ is defined as
\begin{align}
 \beta_{\delta}(p_X\|q_X) \triangleq
\inf\limits_{\substack{r_{Z|X} \in \mathcal{Q}(\{0,1\}|\mathcal{X}): \\ \int_{\mathcal{X}}r_{Z|X}(1|x)p_X(x)\, \mathrm{d}x\ge \delta}} \int_{\mathcal{X}}r_{Z|X}(1|x)q_X(x)\, \mathrm{d}x.\label{eqDefISDivergence}
\end{align}
\end{Definition}
\medskip
The existence of a minimizing test $r_{Z|X}$ is guaranteed by the Neyman-Pearson lemma, so the $\inf$ in \eqref{eqDefISDivergence} can be replaced by $\min$. We state in the following lemma and proposition some important properties of $\beta_{\delta}(p_X\|q_X)$, which are crucial for the proof of Theorem~\ref{thmMainResult}. The proof of the two statements in the following lemma can be found in~\cite[Lemma~1]{wang09} and~\cite[Sec.~2.3]{Pol10} respectively.
\medskip
\begin{Lemma}\label{lemmaDPI} Let $p_{X}$ and $q_{X}$ be two probability distributions on some $\mathcal{X}$, and let $g$ be a function whose domain contains $\mathcal{X}$. Then, the following two statements hold:
\begin{enumerate}
\item[1.] (Data processing inequality (DPI)) $\beta_{\delta}(p_X\|q_X) \le  \beta_{\delta}(p_{g(X)}\|q_{g(X)})$.
\item[2.] For all $\xi>0$, $\beta_{\delta}(p_X\|q_X)\ge \frac{1}{\xi}\left(\delta - \int_{\mathcal{X}}p_X(x) \mathbf{1}\left\{ \frac{p_X(x)}{q_X(x)} \ge \xi \right\}\mathrm{d}x\right)$.
\end{enumerate}
\end{Lemma}
\medskip
The proof of the following proposition is similar to Lemma~3 in \cite{wang09} and therefore omitted.
\medskip
\begin{Proposition} \label{propositionBHTLowerBound}
Let $p_{U,V}$ and $s_V$ be two probability distributions defined on $\mathcal{W}\times \mathcal{W}$ and $\mathcal{W}$ respectively for some $\mathcal{W}$, and let $p_U$ be the marginal distributions of $p_{U,V}$. Suppose $p_{U}$ is the uniform distribution, and let
\begin{equation*}
\varepsilon = \Pr\{U\ne V\} 
\end{equation*}
be a real number in $[0, 1)$. Then,
\begin{equation*}
\beta_{1-\varepsilon}(p_{U,V}\|p_{U} s_V) \le \frac{1}{|\mathcal{W}|}. 
\end{equation*}
\end{Proposition}
\section{Achievability Proof of Theorem~\ref{thmMainResult}}\label{secAchievability}
In this section, we will show that for all $\varepsilon\in(0,1)$
\begin{equation}
C_\varepsilon \ge  \max\limits_{0\le\alpha\le 1}R_{\text{cut-set}}\left(\alpha, \frac{P_1}{1-\varepsilon}, \frac{P_2}{1-\varepsilon}\right). \label{achProofSt}
\end{equation}
To this end, we fix $\varepsilon\in(0,1)$ and define $\tilde \alpha\in(0, 1]$ as follows: If $\frac{P_1}{N_2}\le \frac{P_1+P_2}{N_2+N_3}$, let $\tilde \alpha \triangleq 1$; otherwise, let
$\tilde \alpha$ be the unique number in $(0,1)$ such that
 \begin{equation*}
\frac{\tilde \alpha P_1}{N_2} = \frac{P_1 + P_2 +2\sqrt{(1-\tilde \alpha)P_1 P_2}}{N_2+N_3}.
 \end{equation*}
The above choice of $\tilde \alpha$ together with the definition of $R_{\text{cut-set}}\left(\alpha, \frac{P_1}{1-\varepsilon}, \frac{P_2}{1-\varepsilon}\right)$ in~\eqref{defRcutset} implies that
 \begin{align}
\mathrm{C}\left(\frac{\tilde \alpha P_1}{(1-\varepsilon)N_2}\right)& =  R_{\text{cut-set}}\left(\tilde \alpha, \frac{P_1}{1-\varepsilon}, \frac{P_2}{1-\varepsilon}\right) \notag\\
& = \max\limits_{0\le\alpha\le 1}R_{\text{cut-set}}\left(\alpha, \frac{P_1}{1-\varepsilon}, \frac{P_2}{1-\varepsilon}\right) \label{alpha*St}
 \end{align}
 and
  \begin{equation}
\frac{\tilde \alpha P_1}{N_2} \le \frac{P_1 + P_2 +2\sqrt{(1-\tilde \alpha)P_1 P_2}}{N_2+N_3}. \label{defAlpha*}
 \end{equation}
Fix a sufficiently large~$n\in\mathbb{N}$ such that
\begin{equation}
39 n^{-1/4} < \varepsilon, \label{sufficientLarge1}
 \end{equation}
 \begin{equation}
 \left(1+\frac{39}{1-\varepsilon}\right)n^{-1/4}\le 1, \label{sufficientLarge1*}
 \end{equation}
 \begin{equation}
n \mathrm{C}\left(\frac{(1-n^{-1/4})\left( \sqrt{(1-\tilde \alpha)P_1} +\sqrt{P_2}\right)^2}{(1-\varepsilon+39n^{-1/4})(N_2+N_3) + (1-n^{-1/4})\tilde \alpha  P_1} \right) - n^{\frac{3}{4}}- 2\log n\ge 0,\label{sufficientLarge2}
 \end{equation}
 and
 \begin{align}
 n\mathrm{C}\left(\frac{(1-n^{-1/4})\tilde \alpha P_1}{ (1-\varepsilon+39n^{-1/4})N_2}\right) - 2n^{3/4} - \frac{13}{4}\log n \ge 0. \label{sufficientLarge3}
 \end{align}
We will construct in Sections~\ref{subsec1} to~\ref{subsec5} an $((L+1)n, M^L, P_1, P_2)$-code, where $L$ and $M$ are two integers which depend on~$n$ and will be specified later in Section~\ref{subsec7}. The corresponding probability of decoding error, the power consumption at the source and the relay, and the rate will be calculated in Sections~\ref{subsec6} and~\ref{subsec7}, Section~\ref{subsecPower}, and Section~\ref{subsecRate} respectively.

\subsection{Message and Submessage Sets} \label{subsec1}
The following strategy of dividing the whole transmission into equal-length blocks is used in the original decode-forward scheme~\cite[Sec.~15.1.4]{CoverBook}. The source transmits information to the destination in $(L+1)n$ channel uses by means of transmitting~$L+1$ blocks of length-$n$ codewords, where each of the first~$L$ blocks carries a new submessage intended for the destination while the last block carries no new submessage. Define the submessage set
\begin{equation*}
\mathcal{W}\triangleq \{1, 2, \ldots, M\}
\end{equation*}
and define the message set
\begin{equation*}
\boldsymbol{\mathcal{W}}\triangleq \overbrace{\mathcal{W}\times \mathcal{W}\times \ldots \times \mathcal{W}}^{L\text{ times}}.
\end{equation*}
Let
\begin{equation*}
\boldsymbol{W}\triangleq (W_1, W_2,\ldots, W_L)
\end{equation*}
be the message intended for the destination where
 $W_\ell\in \mathcal{W}$ is the $\ell^{\text{th}}$ submessage chosen to be transmitted in the $\ell^{\text{th}}$ block for each $\ell\in\{1,2, \ldots, L\}$. The message $\boldsymbol{W}$ is uniformly chosen from $\boldsymbol{\mathcal{W}}$, which implies that the submessage $W_\ell$ is uniform on $\mathcal{W}$ for each $\ell\in\{1, 2, \ldots, L\}$.

\subsection{Generations of Random Codebooks and Random Binning Function} \label{subset2}
Similar to the original decode-forward scheme described in~\cite[Sec.~15.1.4]{CoverBook}, we will construct a random binning function used by both the source and the relay, a random codebook used by the source, and a random codebook used by the relay. Construct an index set for binning
 \begin{equation*}
\mathcal{B} \triangleq \{1, 2, \ldots, B\}
\end{equation*}
where $B$ depends on~$n$ and will be specified later, and construct a random binning function $g:\mathcal{W}\rightarrow \mathcal{B}$ such that
\begin{equation}
\Pr\{g(w)=b\}=\frac{1}{B} \label{labelBinningFunction}
 \end{equation}
 for each $w\in \mathcal{W}$ and each $b\in \mathcal{B}$ where the randomness of~$g$ is not explicitly specified for notational simplicity. Define \begin{equation}
 p_{U^n}\triangleq \mathcal{N}(u^n; 0, 1-n^{-1/4}) \label{defDistU}
  \end{equation}
  and
  \begin{equation}
  p_{V^n}\triangleq \mathcal{N}(v^n; 0, 1-n^{-1/4}), \label{defDistV}
   \end{equation}
   and define $p_U \triangleq p_{U_1}$ and $p_V\triangleq p_{V_1}$. Construct two sets of independently generated codewords $\{U^n(w)\in\mathbb{R}^n|w\in \mathcal{W}\}$ and $\{V^n(b)\in\mathbb{R}^n|b\in \mathcal{B}\}$, each consisting of i.i.d.\ codewords such that $U^n(w) \sim p_{U^n}$ for each $w\in\mathcal{W}$ and $V^n(b)\sim p_{V^n}$ for each $b\in\mathcal{B}$. The variances of $U$ and $V$ have been chosen to be slightly less than~$1$ so that
 \begin{align}
   \Pr_{p_{U^n}}\{\|U^n\|^2>n\}
&\stackrel{\text{(a)}}{\le} 
 \frac{2n(1-n^{-1/4})^2}{n^{3/2}}\notag \\
&\le \frac{2}{n^{1/2}} \label{powerOutage*}
\end{align}
where (a) follows from~\eqref{defDistU} and Chebyshev's inequality. For each $w\in\mathcal{W}$ and each $b\in \mathcal{B}$, define the random codeword
 \begin{align}
 \tilde X_1^n(w,b) \triangleq \sqrt{\frac{\tilde \alpha P_1}{1-\varepsilon+39 n^{-1/4}}}\, U^n(w) + \sqrt{\frac{(1-\tilde \alpha) P_1}{1-\varepsilon+39 n^{-1/4}}}\, V^n(b). \label{defTildeX1}
 \end{align}
 In addition, define for each $w\in\mathcal{W}$ and each $b\in \mathcal{B}$ the random codeword
  \begin{align}
 \tilde X_2^n(b) \triangleq \sqrt{\frac{P_2}{1-\varepsilon+39 n^{-1/4}}}\, V^n(b).  \label{defTildeX2}
 \end{align}
 To facilitate discussion in the following subsections, define
 \begin{align}
 P_i^{(n)}\triangleq \frac{P_i}{1-\varepsilon + 39 n^{-1/4}} \label{defPin}
 \end{align}
 for each $i\in\{1,2\}$, define
 \begin{align}
p_{\tilde X_1, \tilde X_2|U, V}(\tilde x_1, \tilde x_2|u, v) \triangleq \mathbf{1}\left\{\tilde x_1= \sqrt{\tilde \alpha P_1^{(n)}} \, u + \sqrt{(1-\tilde \alpha) P_1^{(n)}}\, v\right\}  \mathbf{1}\left\{\tilde x_2= \sqrt{P_2^{(n)}} \, v \right\} \label{defDistForDecodingBefore}
 \end{align}
 for all $(u, v, \tilde x_1, \tilde x_2)\in \mathbb{R}^4$ and define
 \begin{equation}
 p_{U,V,\tilde X_1,\tilde X_2, Y_2, Y_3}\triangleq p_U p_V p_{\tilde X_1, \tilde X_2|U, V} q_{Y_2, Y_3|\tilde X_1, \tilde X_2}. \label{defDistForDecoding}
 \end{equation}

\subsection{Superposition Coding with Power Control at the Source} \label{subsec3}
In this subsection, we describe a power control strategy used by the source as suggested in~\cite{YCDP15,TFT15} as well as a superposition coding scheme as in the original decode-forward scheme~\cite[Sec.~15.1.4]{CoverBook}. Recall $\varepsilon > 39 n^{-1/4}$ by \eqref{sufficientLarge1} and partition $\mathcal{W}$ into two sets
\begin{equation}
\mathcal{A}\triangleq \left\{1, 2, \ldots, \left\lceil \left(1-\varepsilon+34 n^{-1/4}\right)M\right\rceil\right\} \label{defSetA}
 \end{equation}
 and $\mathcal{A}^c\triangleq \mathcal{W}\setminus \mathcal{A}$.
Consider the following superposition coding strategy combined with power control. To send message~$\boldsymbol{W}$, the source
transmits in each block~$\ell\in\{1, 2, \ldots, L+1\}$
\begin{align}
X_1^n(\ell)\triangleq
\begin{cases}
\tilde X_1^n(W_1, g(W_0)) & \text{if $\ell=1$ and $\|\tilde X_1^n(W_1, g(W_0))\|^2 \le nP_1^{(n)}$,}\\
\tilde X_1^n(W_\ell, g(W_{\ell-1})) & \text{if $\ell\ge 2$, $\|\tilde X_1^n(W_\ell, g(W_{\ell-1}))\|^2 \le nP_1^{(n)}$ and $W_1\in \mathcal{A}$,}\\ 0^{n} & \text{otherwise,}
\end{cases} \label{sourceTransmissionRule}
\end{align}
 where we use the convention that
 \begin{equation}
 W_0=W_{L+1}= 1 \label{conventionDet}
 \end{equation}
 deterministically.
It follows from~\eqref{sourceTransmissionRule} that
\begin{align*}
X_1^{(L+1)n}=(X_1^n(1), X_1^n(2), \ldots, X_1^n(L+1)).
\end{align*}
The power control strategy at the source is captured by the novel transmission rule~\eqref{sourceTransmissionRule}, which prescribes the source to remain silent starting from block~$2$ if $W_1\in \mathcal{A}^c$ (occurs with probability close to $\varepsilon - 34 n^{-1/4}$ by~\eqref{defSetA}).

\subsection{Decode-Forward at the Relay} \label{subsec4}
This subsection describes the decoding and binning strategies performed by the relay under the original decode-forward scheme~\cite[Sec.~15.1.4]{CoverBook}, where the typicality decoding strategy in the original decode-forward scheme is replaced by the Shannon's threshold decoding strategy~\cite{sha57}.
 Let $Y_2^n(\ell)$ denote the symbols received by the relay in block~$\ell$ for each $\ell\in\{1, 2, \ldots, L+1\}$ such that
\begin{equation*}
Y_2^{(L+1)n} = \left(Y_2^n(1), Y_2^n(2), \ldots, Y_2^n(L+1)\right).
\end{equation*}
For each $\ell\in\{1, 2, \ldots, L\}$, let $W_\ell^*$ be the estimate of $W_\ell$ output by the relay. We use the convention that
\begin{equation}
W_0^*\triangleq W_{0} \stackrel{\eqref{conventionDet}}{=}1 \label{conventionW*}
 \end{equation}
 deterministically. Upon receiving $Y_2^n(\ell)$ and having generated $W_1^*, W_2^*, \ldots, W_{\ell-1}^*$, the relay claims that $W_\ell^* = w_\ell^*\in \mathcal{W}$ is the transmitted submessage $W_\ell$ carried by block~$\ell$ if $w_\ell^*$ is the unique integer in $\mathcal{W}$ such that
\begin{align}
(U^n(w_\ell^*), V^n(g(W_{\ell-1}^*)), Y_2^n(\ell)) \in \mathcal{T}_{p_{U, V, Y_2}}^{(n)}(U; Y_2|V) \label{thresholdDecodingRelay}
\end{align}
where the distribution~$p$ was defined in~\eqref{defDistForDecoding}; if no such unique $w_\ell^*$ exists, the relay lets $W_\ell^*$ be uniformly distributed on~$\mathcal{W}$. 
For each $\ell\in\{1, 2, \ldots, L+1\}$, the relay transmits in block~$\ell$
\begin{align}
X_2^n(\ell) \triangleq
\begin{cases}
0^n & \text{if $\ell=1$,}\vspace{0.04 in}\\
\tilde X_2^n(g(W_{\ell-1}^*))  &  \parbox[c]{3.2 in}{if $\ell\ge 2$, $\|\tilde X_2^n(g(W_{\ell-1}^*)) \|^2 \le n P_2^{(n)}$ and $W_1^*\in \mathcal{A}$,}\vspace{0.04 in}\\
0^n & \text{otherwise}.
\end{cases} \label{relayDecodeForward}
\end{align}
It follows from~\eqref{relayDecodeForward} that
\begin{align*}
X_2^{(L+1)n} &= \left(X_2^n(1), X_2^n(2), X_2^n(3), \ldots, X_2^n(L+1)\right)\notag\\
& = \left(0^n, X_2^n(2), X_2^n(3), \ldots, X_2^n(L+1)\right).
\end{align*}
The power control strategy at the relay is captured by the novel transmission rule~\eqref{relayDecodeForward}, which prescribes the relay to remain silent starting from block~$2$ if $W_1^*\in \mathcal{A}^c$ (occurs with probability more than $\varepsilon-39 n^{-1/4}$ by~\eqref{averagePowerEq5}).
\subsection{Sliding Window Decoding at the Destination} \label{subsec5}
The destination uses the sliding window decoding strategy proposed in
\cite[Sec.~IV]{KGG05} where the typicality decoding strategy in the original decode-forward scheme is replaced by Shannon's threshold decoding (of the information density) strategy~\cite{sha57}. Note that if the sliding window decoding strategy is replaced by the backward decoding strategy in the original decode-forward scheme~\cite[Sec.~15.1.4]{CoverBook}, the same non-asymptotic lower bound on the coding rate as shown in~\eqref{achRateEq8} will result. Let~$Y_3^n(\ell)$ denote the length-$n$ codeword in block~$\ell$ received by the destination for each $\ell\in\{1, 2, \ldots, L+1\}$ such that
\begin{align*}
Y_3^{(L+1)n}\triangleq \left(Y_3^n(1), Y_3^n(2), \ldots, Y_3^n(L+1)\right).
\end{align*}
The destination will produce estimates of the submessages according to this order $\hat W_1, \hat W_2, \ldots, \hat W_L$ in a recursive manner described below, where $\hat W_\ell$ denote the estimate of~$W_\ell$ and we use the convention that
\begin{align*}
\hat W_0\triangleq W_0\stackrel{\eqref{conventionDet}}{=}1
\end{align*}
deterministically. For each $\ell\in\{1, 2, \ldots, L\}$, given that $Y_3(\ell)$ and $Y_3(\ell+1)$ have been received and $\{\hat W_h\}_{h=0}^{\ell-1}$ have been produced, the destination will output $\hat W_\ell$ according to the following two cases:\\
\textbf{Case $\ell=1$\,:}
First, the destination decodes the bin index of $W_1$ based on $Y_1^n(2)$ by using the following threshold decoding rule: Let $\hat B_1$ denote the estimate of $g(W_1)$. The destination claims that $\hat B_1=\hat b_1\in \mathcal{B}$ is the bin index of the transmitted $W_1$ if $\hat b_1$ is the unique integer in~$\mathcal{B}$ that satisfies
\begin{align}
(V^n(\hat b_1), Y_3^n(2))\in \mathcal{T}_{p_{V,Y_3}}^{(n)}(V;Y_3) \label{binDecodingDestinationCase2}
\end{align}
where the distribution~$p$ was defined in~\eqref{defDistForDecoding};
if no such unique $\hat b_1$ exists, the destination lets $\hat B_1$ be uniformly distributed on~$\mathcal{B}$. Then, the destination claims that $\hat W_1 = \hat w_1\in g^{-1}(\hat B_1)$ is the transmitted submessage if $\hat w_1$ is the unique integer in $g^{-1}(\hat B_1)$ such that
\begin{align}
(U^n(\hat w_1), V^n(g(1)), Y_3^n(1)+V^n(g(1))) \in \mathcal{T}_{p_{U,V,Y_3}}^{(n)}(U;Y_3|V);\label{listDecodingAtDestinationCase2}
\end{align}
if no such unique $\hat w_1$ exists, the destination lets $\hat W_1$ be uniformly distributed on~$\mathcal{W}$.
Note that since the relay transmits nothing in the first block, $Y_3^n(1)+V^n(g(1))$ is distributed according to the $n$-fold product of $p_{Y_3}$ defined in~\eqref{defDistForDecoding} and hence the decoding rule in~\eqref{listDecodingAtDestinationCase2} makes sense.\\
\textbf{Case $\ell\ge 2$\,:}
First, the destination decodes the bin index of $W_\ell$ based on $Y_3^n(\ell+1)$ by using the following threshold decoding rule: Let $\hat B_\ell$ denote the estimate of $g(W_\ell)$. The destination claims that $\hat B_\ell=\hat b_\ell\in \mathcal{B}$ is the bin index of the transmitted $W_\ell$ if $\hat b_\ell$ is the unique integer in $\mathcal{B}$ that satisfies
\begin{align}
(V^n(\hat b_\ell), Y_3^n(\ell+1)) \in \mathcal{T}_{p_{V,Y_3}}^{(n)}(V;Y_3); \label{binDecodingDestination}
\end{align}
if no such unique $\hat b_\ell$ exists, the destination lets $\hat B_\ell$ be uniformly distributed on~$\mathcal{B}$. Then, the destination claims that $\hat W_\ell = \hat w_\ell\in g^{-1}(\hat B_\ell)$ is the transmitted submessage if $\hat w_\ell$ is the unique integer in $g^{-1}(\hat B_\ell)$ such that
\begin{align}
(U^n(\hat w_\ell), V^n(g(\hat W_{\ell-1})), Y_3^n(\ell))\in \mathcal{T}_{p_{U,V,Y_3}}^{(n)}(U;Y_3|V);  \label{listDecodingAtDestination}
\end{align}
if no such unique $\hat w_\ell$ exists, the destination lets $\hat W_\ell$ be uniformly distributed on~$\mathcal{W}$.

\subsection{Calculation of Error Probability} \label{subsec6}
Let $r_{\boldsymbol{W}, g(\boldsymbol{W}), \tilde{\boldsymbol{X}}_1,  \tilde{\boldsymbol{X}}_2, \boldsymbol{X}_1, \boldsymbol{X}_2, \boldsymbol{Y}_2, \boldsymbol{Y}_3, \boldsymbol{W}^*, \boldsymbol{\hat B}, \hat{\boldsymbol{W}}}$ be the distribution induced by the random coding scheme described from Sections~\ref{subsec1} to~\ref{subsec5}, where
\begin{equation*}
g(\boldsymbol{W})\triangleq (g(W_\ell)|\ell\in\{0, 1, \ldots, L+1\}),
\end{equation*}
 \begin{equation*}
 \tilde{\boldsymbol{X}}_1\triangleq (\tilde X_1^n(W_\ell, g(W_{\ell-1}))|\ell\in\{1,2, \ldots, L+1\}),
  \end{equation*}
  \begin{equation*}
  \tilde{\boldsymbol{X}}_2\triangleq (\tilde X_2^n(g(W_{\ell-1}^*))|\ell\in\{1, 2, \ldots, L+1\}),
    \end{equation*}
    \begin{equation*}
    \boldsymbol{X}_1\triangleq X_1^{(L+1)n},
  \end{equation*}
     \begin{equation*}
     \boldsymbol{X}_2\triangleq X_2^{(L+1)n},
       \end{equation*}
       \begin{equation*}
       \boldsymbol{Y}_2\triangleq Y_2^{(L+1)n},
          \end{equation*}
            \begin{equation*}
            \boldsymbol{Y}_3\triangleq Y_3^{(L+1)n},
          \end{equation*}
            \begin{equation*}
            \boldsymbol{W}^*\triangleq (W_\ell^*|\ell\in\{1, 2, \ldots, L\}),
               \end{equation*}
                 \begin{equation*}
                 \boldsymbol{\hat B}\triangleq (\hat B_\ell|\ell\in\{1, 2, \ldots, L\})
                    \end{equation*}
                    and
                       \begin{equation*}
                       \hat{\boldsymbol{W}}\triangleq (\hat W_\ell|\ell\in\{1, 2, \ldots, L\}).
                        \end{equation*}
                        To simplify notation, we omit the subscripts of the probability and expectation terms which are evaluated according to~$r$ in the rest of Section~\ref{secAchievability}. We are interested in bounding
\begin{align}
&\Pr\left\{\boldsymbol{W} \ne \hat{\boldsymbol{W}}\right\}\notag\\
&= \Pr\left\{\{\hat{\boldsymbol{W}}\ne \boldsymbol{W}\}\cap \{W_1\in \mathcal{A}\}\cap\{\boldsymbol{W}^* = \boldsymbol{W}\} \right\}  +  \Pr\left\{\{\hat{\boldsymbol{W}}\ne \boldsymbol{W}\}\cap \left( \{W_1\in \mathcal{A}^c\}\cup\{\boldsymbol{W}^* \ne \boldsymbol{W}\}\right) \right\}\notag\\
&\le \Pr\left\{\{\hat{\boldsymbol{W}}\ne \boldsymbol{W}\}\cap \{W_1\in \mathcal{A}\}\cap\{\boldsymbol{W}^* = \boldsymbol{W}\} \right\}  +  \Pr\left\{\{W_1\in \mathcal{A}^c\}\cup\{\boldsymbol{W}^* \ne \boldsymbol{W}\} \right\}\notag\\
&\stackrel{\text{(a)}}{\le} \Pr\left\{\{\hat{\boldsymbol{W}}\ne \boldsymbol{W}\}\cap \{W_1\in \mathcal{A}\}\cap\{\boldsymbol{W}^* = \boldsymbol{W}\} \right\} +   \varepsilon - 34 n^{-1/4} \notag\\
&\qquad +\sum_{\ell=1}^L  \Pr\left\{\{W_1\in \mathcal{A}\}\cap\{W_\ell^*\ne W_\ell\}\cap \bigcap_{h=1}^{\ell-1}\{W_h^*= W_h\}\right\} \label{errorProbCalEq1}
\end{align}
where (a) follows from the union bound and the fact due to~\eqref{defSetA} that
\begin{align*}
\Pr\left\{ W_1\in \mathcal{A}^c \right\} \le  \varepsilon - 34 n^{-1/4}. 
\end{align*}
In Section~\ref{subsubsec1} to follow, we will obtain an upper bound on $\Pr\left\{\{W_1\in \mathcal{A}\}\cap\{W_\ell^*\ne W_\ell\}\cap\bigcap_{h=1}^{\ell-1}\{W_h^*= W_h\}\right\}$ for each $\ell\in \{1, 2, \ldots, L\}$, which characterizes the decoding error probability  of the submessage~$W_\ell$ at the relay. In Sections~\ref{subsubsec2} and~\ref{subsubsec3}, we will obtain an upper bound on $\Pr\left\{\{\hat{\boldsymbol{W}}\ne \boldsymbol{W}\}\cap \{W_1\in \mathcal{A}\}\cap\{\boldsymbol{W}^* = \boldsymbol{W}\} \right\}$, which characterizes the error probability of decoding the overall message $\boldsymbol{W}$ at the destination.
\subsubsection{Error Probabilities of Decoding Submessages at the Relay} \label{subsubsec1}
To simplify notation, for each block $\ell\in\{1, 2, \ldots, L+1\}$, we let
\begin{equation}
\mathcal{E}_{1, \ell}\triangleq \{\|\tilde X_1^n(W_\ell, g(W_{\ell-1}))\|^2 \le nP_1^{(n)}\} \label{defEventE1ell}
\end{equation}
and
\begin{equation}
\mathcal{E}_{2, \ell}\triangleq \{\|\tilde X_2^n(g(W_{\ell-1}))\|^2 \le nP_2^{(n)}\} \label{defEventE2ell}
\end{equation}
denote the events that the codewords $\tilde X_1^n(W_\ell, g(W_{\ell-1}))$ and~$\tilde X_2^n(g(W_{\ell-1}))$ satisfy the respective peak power constraints. For each $i\in\{1,2\}$ and each $\ell\in\{1, 2, \ldots, L+1\}$, we have the following due to the definitions of $\mathcal{E}_{1,\ell}$ and $\mathcal{E}_{2, \ell}$ in~\eqref{defEventE1ell} and~\eqref{defEventE2ell} respectively, the definitions of $p_{U^n}$ and $p_{V^n}$ in \eqref{defDistU} and~\eqref{defDistV} respectively, the definitions of $\tilde X_1^n(W_\ell, g(W_{\ell-1}))$ and $\tilde X_2^n(g(W_{\ell-1}))$ in~\eqref{defTildeX1} and~$\eqref{defTildeX2}$ respectively, and the definition of~$P_i^{(n)}$ in~\eqref{defPin}:
\begin{align*}
\Pr\left\{\mathcal{E}_{i, \ell}^c\right\}
& = \Pr_{p_{\tilde X_i^n}}\{\|\tilde X_i^n\|^2>nP_i^{(n)}\}\\
&= \Pr_{p_{U^n}}\{\|U^n\|^2>n\},
\end{align*}
which implies from~\eqref{powerOutage*} that
\begin{equation}
\Pr\left\{\mathcal{E}_{i, \ell}^c\right\} \le \frac{2}{n^{1/2}}. \label{powerOutage}
\end{equation}
Let
\begin{align}
\mathcal{G}_1\triangleq \{W_1\in \mathcal{A}\} \label{defEventG1}
\end{align}
be the event that $W_1$ is a ``good" message that falls inside~$\mathcal{A}$ so that the source will keep transmitting information beyond the first block according to~\eqref{sourceTransmissionRule}, and let
\begin{align}
\mathcal{G}_2\triangleq \{W_1^*\in \mathcal{A}\} \label{defEventG2}
\end{align}
be the event that the relay's estimate $W_1^*$ is a ``good" message that falls inside~$\mathcal{A}$ so that the relay will keep transmitting information beyond the first block according to~\eqref{relayDecodeForward}.
In addition, for each block $\ell\in\{1, 2, \ldots, L+1\}$, we let
\begin{align}
\mathcal{F}_{1, \ell} \triangleq \{X_1^n(\ell) = \tilde X_1^n(W_\ell, g(W_{\ell-1}))\} \label{defEventF1}
\end{align}
and
\begin{align}
\mathcal{F}_{2, \ell} \triangleq
\begin{cases}\{X_2^n(\ell) = 0^n\} & \text{if $\ell=1$,}\\ \{X_2^n(\ell) = \tilde X_2^n(g(W_{\ell-1}))\} & \text{otherwise}\end{cases} \label{defEventF2}
\end{align}
be the respective events that the first two cases of the transmission rule of the source in~\eqref{sourceTransmissionRule} and the transmission rule of the relay in~\eqref{relayDecodeForward} occur, let
\begin{align}
\mathcal{H}_{\ell-1} \triangleq \{W_{\ell-1}^* = W_{\ell-1}\} \label{defEventH}
\end{align}
be the event that the relay correctly decodes $W_{\ell-1}$, let
\begin{align}
\mathcal{I}_\ell \triangleq \{\hat B_\ell = g(W_\ell)\} \label{defEventI1}
\end{align}
be the event that the destination correctly decodes the bin index of $W_\ell$, and let
\begin{align}
\mathcal{J}_\ell \triangleq \{\hat W_\ell = W_\ell\} \label{defEventI2}
\end{align}
be the event that the destination correctly decodes $W_\ell$. To facilitate understanding, the descriptions of the nine previously defined events are listed in Table~\ref{table1}.
\begin{table}[t]
 \centering
  \caption{Descriptions of Events for $1\le \ell \le L+1$.}
  \label{table1}
  \begin{tabular}{ccc}
    \toprule
  Event & Description & Equation\\
    \midrule
   $\mathcal{E}_{1, \ell}$ & The random codeword formed by the source satisfies the peak power constraint with admissible power $P_1^{(n)}$  & \eqref{defEventE1ell}\\
    $\mathcal{E}_{2, \ell}$ &  The random codeword formed by the relay satisfies the peak power constraint with admissible power $P_2^{(n)}$ & \eqref{defEventE2ell}\\
        $\mathcal{F}_{1, \ell}$ & The codeword transmitted by the source is equal to the randomly generated codeword & \eqref{defEventF1}\\
         $\mathcal{F}_{2, \ell}$ &  The codeword transmitted by the relay is equal to the randomly generated codeword for $\ell\ge 2$ and $0^n$ for $\ell=1$ & \eqref{defEventF2}\\
  $\mathcal{G}_1$ & The submessage $W_1$ is a ``good" message that falls inside~$\mathcal{A}$ & \eqref{defEventG1} \\
   $\mathcal{G}_2$ &  The relay's estimate $W_1^*$ is a ``good" message that falls inside~$\mathcal{A}$ & \eqref{defEventG2} \\
    $\mathcal{H}_{\ell-1}$ & The relay correctly decodes the submessage $W_{\ell-1}$ & \eqref{defEventH} \\
    $\mathcal{I}_\ell$ & The destination correctly decodes the bin index of the submessage $W_{\ell}$ & \eqref{defEventI1} \\
   $\mathcal{J}_\ell$ &  The destination correctly decodes the submessage $W_{\ell}$ & \eqref{defEventI2} \\
    \bottomrule
  \end{tabular}
\end{table}
Following~\eqref{errorProbCalEq1},
we consider the following chain of inequalities for each $\ell\in\{1, 2, \ldots, L\}$:
\begin{align}
\Pr\left\{\mathcal{G}_1\cap\mathcal{H}_{\ell}^c\cap\bigcap_{h=1}^{\ell-1}\mathcal{H}_h\right\}
&\stackrel{\text{(a)}}{\le} \Pr\left\{\mathcal{G}_1\cap\mathcal{H}_{\ell}^c\cap\bigcap_{h=1}^{\ell-1}\mathcal{H}_h\cap \mathcal{E}_{1, \ell}\right\} \notag + \Pr\left\{\mathcal{E}_{1, \ell}^c\right\} \notag\\
& \stackrel{\eqref{powerOutage}}{\le} \Pr\left\{\mathcal{G}_1\cap\mathcal{H}_{\ell}^c\cap\bigcap_{h=1}^{\ell-1}\mathcal{H}_h\cap \mathcal{E}_{1, \ell}\right\} +\frac{2}{n^{1/2}}  \notag\\
& \le  \Pr\left\{\mathcal{H}_{\ell}^c\cap \mathcal{E}_{1, \ell} \cap\mathcal{G}_1\cap \mathcal{H}_{\ell-1}\right\}+\frac{2}{n^{1/2}}\notag\\*
& \stackrel{\eqref{sourceTransmissionRule}}{\le}  \Pr\left\{\left.\mathcal{H}_{\ell}^c\cap \mathcal{F}_{1, \ell} \right|\mathcal{H}_{\ell-1}\right\}+\frac{2}{n^{1/2}} \label{errorProbCalEq3Temp1*}
\end{align}
where (a) follows from the union bound.
In order to simplify the first term in~\eqref{errorProbCalEq3Temp1*}, we recall
the definition of $p_{U,V,Y_2,Y_3}$ in~\eqref{defDistForDecoding} and define
\begin{align*}
p_{U^n, V^n, Y_2^n,Y_3^n}(u^n, v^n, y_2^n,y_3^n)\triangleq \prod_{k=1}^n p_{U, V, Y_2, Y_3}(u_k, v_k, y_{2,k}, y_{3,k}) 
\end{align*}
for each $(u^n, v^n, y_2^n, y_3^n)\in\mathbb{R}^n \times \mathbb{R}^n \times \mathbb{R}^n \times \mathbb{R}^n$. In addition, we let
\begin{equation}
p_{U^n(i)}(u^n)\triangleq p_{U^n}(u^n) \label{defDistUn}
 \end{equation}
 for each $i\in\{1, 2, \ldots, M\}$, let
 \begin{equation}
 p_{V^n(b)}(v^n)\triangleq p_{V^n}(v^n) \label{defDistVn}
  \end{equation}
  for each $b\in\{1, 2, \ldots, B\}$, and let
  \begin{align}
  p_{U^n(1), V^n(1),Y_2^n, Y_3^n}(u^n, v^n, y_2^n, y_3^n)\triangleq p_{U^n(1)}(u^n)p_{V^n(1)}(v^n) p_{Y_2^n, Y_3^n|U^n, V^n}(y_2^n, y_3^n|u^n, v^n). \label{defDistUVYn}
   \end{align}
   By the construction rule of~$\tilde X_1^n(W_\ell, g(W_{\ell-1}))$ in~\eqref{defTildeX1} and the threshold decoding rule~\eqref{thresholdDecodingRelay} used by the relay, the first term in~\eqref{errorProbCalEq3Temp1*} can be bounded as
\begin{align}
&\Pr\left\{\left.\mathcal{H}_{\ell}^c\cap \mathcal{F}_{1, \ell} \right|\mathcal{H}_{\ell-1}\right\}\notag\\
 & \le \Pr_{p_{U^n(1), V^n(1), Y_2^n}}\left\{(U^n(1), V^n(1), Y_2^n)\notin \mathcal{T}_{p_{U,V,Y_2}}^{(n)}(U;Y_2|V)\right\} \notag\\
&\qquad + \Pr_{\left(\prod_{i=1}^M p_{U^n(i)}\right)p_{V^n(1)}p_{Y_2^n|U^n(1), V^n(1)} }\left\{\bigcup_{j=2}^M \left\{(U^n(j), V^n(1), Y_2^n) \in \mathcal{T}_{p_{U,V,Y_2}}^{(n)}(U;Y_2|V)\right\}\right\} \notag\\*
&\stackrel{\text{(a)}}{\le} \frac{1}{n^{1/2}}+ (M-1)   e^{-\left( n\E\left[\log\left(\frac{p(Y_2|U,V)}{p(Y_2|V)}\right)\right] - \sqrt{n^{\frac{3}{2}}\Var\left[\log\left(\frac{p(Y_2|U,V)}{p(Y_2|V)}\right)\right]}\right)} \label{errorProbCalEq3Temp1}
\end{align}
where (a) follows from Lemma~\ref{lemmaPacking}.
\subsubsection{Error Probabilities of Decoding Bin Indices at the Destination} \label{subsubsec2}
First, we follow~\eqref{errorProbCalEq1} and consider the following quantity related to the error probability of decoding the overall message at the destination:
\begin{align}
&\Pr\left\{\{\hat{\boldsymbol{W}}\ne \boldsymbol{W}\}\cap \mathcal{G}_1\cap\{\boldsymbol{W}^* = \boldsymbol{W}\} \right\}\notag\\
&\stackrel{\text{(a)}}{\le} \sum_{\ell=1}^L \Pr\left\{\mathcal{J}_\ell^c \cap \bigcap_{h=1}^{\ell}\mathcal{J}_{h-1}\cap \mathcal{G}_1\cap\{\boldsymbol{W}^* = \boldsymbol{W}\} \right\}\notag\\
& \le  \sum_{\ell=1}^L  \Pr\left\{\mathcal{J}_\ell^c \cap \mathcal{J}_{\ell-1}\cap \mathcal{G}_1 \cap \mathcal{G}_2 \cap \mathcal{H}_{\ell-1} \cap \mathcal{H}_\ell \right\} \notag\\
& \stackrel{\text{(b)}}{\le}  \sum_{\ell=1}^L  \Big(\Pr\left\{\mathcal{J}_\ell^c \cap \mathcal{J}_{\ell-1}\cap \mathcal{I}_\ell\cap\mathcal{G}_1\cap \mathcal{G}_2 \cap  \mathcal{H}_{\ell-1} \right\}    + \Pr\left\{\mathcal{I}_\ell^c \cap \mathcal{G}_1\cap \mathcal{G}_2\cap \mathcal{H}_\ell \right\}\Big) \label{errorProbCalEq5}
\end{align}
where (a) and (b) follow from the union bound. The second term in the summation in~\eqref{errorProbCalEq5} characterizes the error probability of decoding the bin index $g(W_\ell)$ at the destination, which is bounded for each $\ell\in\{1, 2, \ldots, L\}$  as follows:
\begin{align}
& \Pr\left\{\mathcal{I}_\ell^c \cap \mathcal{G}_1\cap \mathcal{G}_2\cap \mathcal{H}_\ell \right\}\notag\\
&\stackrel{\text{(a)}}{\le} \Pr\left\{\mathcal{I}_\ell^c \cap \mathcal{G}_1\cap \mathcal{G}_2\cap \mathcal{H}_\ell\cap \mathcal{E}_{1, \ell+1}\cap \mathcal{E}_{2,\ell+1} \right\} + \Pr\{\mathcal{E}_{1, \ell+1}^c\} + \Pr\{\mathcal{E}_{2, \ell+1}^c\}\notag\\
&\stackrel{\eqref{powerOutage}}{\le} \Pr\left\{\mathcal{I}_\ell^c \cap \mathcal{G}_1\cap \mathcal{G}_2\cap \mathcal{H}_\ell\cap \mathcal{E}_{1, \ell+1}\cap \mathcal{E}_{2,\ell+1} \right\}   + \frac{4}{n^{1/2}}\notag\\
&\stackrel{\text{(b)}}{\le} \Pr\left\{ \mathcal{I}_\ell^c \cap \mathcal{F}_{1, \ell+1} \cap \mathcal{F}_{2, \ell+1} \right\} + \frac{4}{n^{1/2}} \notag\\
&\stackrel{\text{(c)}}{\le}  \frac{4}{n^{1/2}} + \Pr_{p_{U^n(1), V^n(1), Y_3^n}}\left\{(V^n(1),Y_3^n) \notin \mathcal{T}_{p_{V,Y_3}}^{(n)}(V;Y_3) \right\} \notag\\*
 &\qquad+ \Pr_{p_{U^n(1)}\left(\prod\limits_{j=1}^B p_{V^n(j)}\right)  p_{Y_3^n|U^n(1), V^n(1)} }\left\{\bigcup_{j=2}^B \left\{(V^n(j),Y_3^n) \in \mathcal{T}_{p_{V,Y_3}}^{(n)}(V;Y_3) \right\} \right\}\notag \\
&\stackrel{\text{(d)}}{\le} \frac{5}{n^{1/2}} + (B-1) e^{-\left( n\E\left[\log\left(\frac{p(Y_3|V)}{p(Y_3)}\right)\right] - \sqrt{n^{\frac{3}{2}}\Var\left[\log\left(\frac{p(Y_3|V)}{p(Y_3)}\right)\right]}\right)} \label{errorProbCalEq6}
\end{align}
where
\begin{enumerate}
\item[(a)] follows from the union bound.
\item[(b)] follows from the transmission rule~\eqref{sourceTransmissionRule} used by the source and the transmission rule~\eqref{relayDecodeForward} used by the relay.
\item[(c)] follows from the threshold decoding rules~\eqref{binDecodingDestination} and~\eqref{binDecodingDestinationCase2} used by the destination for obtaining $\hat B_\ell$, the construction rules of~$\tilde X_1^n(W_{\ell+1}, g(W_\ell))$ and~$\tilde X_2^n(g(W_\ell))$ in~\eqref{defTildeX1} and~\eqref{defTildeX2} respectively, and the definitions of $p_{U^n(i)}$, $p_{V^n(b)}$ and $p_{U^n(1), V^n(1), Y_3^n}$ in~\eqref{defDistUn}, \eqref{defDistVn} and~\eqref{defDistUVYn} respectively.
   \item[(d)] follows from Lemma~\ref{lemmaPacking}.
\end{enumerate}
\subsubsection{Error Probabilities of Decoding Submessages at the Destination} \label{subsubsec3}
The first term in the summation in~\eqref{errorProbCalEq5} characterizes the error probability of decoding the submessage $W_\ell$ at the destination, which is bounded for each $\ell\in\{1, 2, \ldots, L\}$  as follows:
\begin{align}
& \Pr\left\{\mathcal{J}_\ell^c  \cap \mathcal{J}_{\ell-1}\cap \mathcal{I}_\ell \cap \mathcal{G}_1\cap \mathcal{G}_2 \cap \mathcal{H}_{\ell-1} \right\}  \notag\\
&\stackrel{\text{(a)}}{\le}  \Pr\left\{\mathcal{J}_\ell^c \cap \mathcal{J}_{\ell-1}\cap \mathcal{I}_\ell \cap \mathcal{G}_1\cap \mathcal{G}_1 \cap \mathcal{H}_{\ell-1}\cap \left\{\left|g^{-1}(\hat B_\ell)\right|\le \frac{nM}{B} \right\}\cap \mathcal{E}_{1, \ell} \cap \mathcal{E}_{2, \ell}\right\}  +\Pr\left\{ \left|g^{-1}(\hat B_\ell)\right|> \frac{nM}{B}  \right\} \notag\\
&\qquad + \Pr\{\mathcal{E}_{1, \ell}^c\} +\Pr\{\mathcal{E}_{2, \ell}^c\}\notag\\
& \stackrel{\text{(b)}}{\le} \Pr\left\{\mathcal{J}_\ell^c\cap \mathcal{J}_{\ell-1} \cap \mathcal{I}_\ell \cap \mathcal{G}_1\cap \mathcal{G}_1 \cap \mathcal{H}_{\ell-1}\cap \left\{\left|g^{-1}(\hat B_\ell)\right|\le \frac{nM}{B} \right\}\cap \mathcal{E}_{1, \ell} \cap \mathcal{E}_{2, \ell}\right\}   + \frac{1}{n} + \frac{4}{n^{1/2}}\notag\\
& \stackrel{\text{(c)}}{\le} \Pr\left\{\mathcal{J}_\ell^c\cap \mathcal{J}_{\ell-1}\cap \left\{\left|g^{-1}(g(W_\ell))\right| \le \frac{nM}{B}  \right\}\cap \mathcal{F}_{1, \ell} \cap \mathcal{F}_{2, \ell} \right\}  + \frac{1}{n} + \frac{4}{n^{1/2}}\notag\\
& \stackrel{\text{(d)}}{\le}  \frac{1}{n}+ \frac{4}{n^{1/2}} +\Pr_{p_{U^n(1), V^n(1), Y_3^n}}\left\{(U^n(1), V^n(1), Y_3^n)\notin \mathcal{T}_{p_{U,V,Y_3}}^{(n)}(U;Y_3|V)\right\}  \notag\\*
& \quad + \Pr_{\left(\prod\limits_{i=1}^{\lfloor\frac{nM}{B}\rfloor} p_{U^n(i)}\right) p_{V^n(1)} p_{Y_3^n|U^n(1), V^n(1)} }\left\{\parbox[c]{2.8 in}{$\bigcup\limits_{j=2}^{\lfloor\frac{nM}{B}\rfloor}\left\{(U^n(j), V^n(1), Y_3^n)\in \mathcal{T}_{p_{U,V,Y_3}}^{(n)}(U;Y_3|V)\right\}$}\right\} \notag\\
&\stackrel{\text{(e)}}{\le}\frac{5}{n^{1/2}}+\frac{1}{n}+ \left(\frac{nM}{B}-1\right)e^{-\left( n\E\left[\log\left(\frac{p(Y_3|U,V)}{p(Y_3|V)}\right)\right] - \sqrt{n^{\frac{3}{2}}\Var\left[\log\left(\frac{p(Y_3|U,V)}{p(Y_3|V)}\right)\right]} \right)}, \label{errorProbCalEq7}
\end{align}
where
\begin{enumerate}
\item[(a)] follows from the union bound.
\item[(b)] follows from \eqref{powerOutage} and the following fact due to Markov's inequality:
 \begin{align*}
 \Pr\left\{ \left|g^{-1}(\hat B_\ell)\right|> \frac{nM}{B}  \right\} &\le \frac{B\E\left[\big|g^{-1}(\hat B_\ell)\big|\right]}{nM}\\
 & \stackrel{\eqref{labelBinningFunction}}{=} \frac{1}{n}.
 \end{align*}
 \item[(c)] follows from the transmission rule~\eqref{sourceTransmissionRule} used by the source, and the transmission rule~\eqref{relayDecodeForward} used by the relay.
\item[(d)] follows from the threshold decoding rules~\eqref{listDecodingAtDestination} and~\eqref{listDecodingAtDestinationCase2} used by the destination for obtaining $\hat W_\ell$, the construction rules of~$\tilde X_1^n(W_{\ell}, g(W_{\ell-1}))$ and~$\tilde X_2^n(g(W_{\ell-1}))$ in~\eqref{defTildeX1} and~\eqref{defTildeX2} respectively, and the definitions of $p_{U^n(i)}$, $p_{V^n(b)}$  and $p_{U^n(1), Y_3^n}$ in~\eqref{defDistUn}, \eqref{defDistVn} and~\eqref{defDistUVYn} respectively.
    \item[(e)] follows from Lemma~\ref{lemmaPacking}.
\end{enumerate}
\subsection{Choices of $B$, $M$ and $L$ for Simplifying the Overall Error Probability}\label{subsec7}
In order to simplify the expectation and variance terms in~\eqref{errorProbCalEq3Temp1}, \eqref{errorProbCalEq6} and~\eqref{errorProbCalEq7}, we set the number of bins to be
\begin{align}
B&\triangleq \Bigg\lceil e^{n\E\left[\log\left(\frac{p(Y_3|V)}{p(Y_3)}\right)\right] - \sqrt{n^{\frac{3}{2}}\Var\left[\log\left(\frac{p(Y_3|V)}{p(Y_3)}\right)\right]}- \log n}\:  \Bigg\rceil \label{defB}
\end{align}
and the number of messages to be
\begin{align}
M&\triangleq \min\left\{\parbox[c]{4.4 in}{$ \Bigg\lceil  e^{n\E\left[\log\left(\frac{p(Y_2|U,V)}{p(Y_2|V)}\right)\right] - \sqrt{n^{\frac{3}{2}}\Var\left[\log\left(\frac{p(Y_2|U,V)}{p(Y_2|V)}\right)\right]}-\log n}\:  \Bigg\rceil , \vspace{0.04 in}\\
\Bigg\lceil e^{ n\E\left[\log\left(\frac{p(Y_3|U, V)}{p(Y_3)}\right)\right] - \sqrt{n^{\frac{3}{2}}\Var\left[\log\left(\frac{p(Y_3|V)}{p(Y_3)}\right)\right]} - \sqrt{n^{\frac{3}{2}}\Var\left[\log\left(\frac{p(Y_3|U,V)}{p(Y_3|V)}\right)\right]} - 3\log n}\:  \Bigg\rceil$}\right\} \label{defM}
\end{align}
After some straightforward calculations based on~\eqref{defChannelInDefinition}, \eqref{defDistU}, \eqref{defDistV}, \eqref{defDistForDecodingBefore} and~\eqref{defDistForDecoding}, we obtain from~\eqref{defB} and~\eqref{defM} that
\begin{align}
B & \ge e^{n \mathrm{C}\left(\frac{(1-n^{-1/4})\left( \sqrt{(1-\tilde \alpha)P_1^{(n)}} +\sqrt{P_2^{(n)}}\right)^2}{N_2+N_3 + (1-n^{-1/4})\tilde \alpha  P_1^{(n)}} \right) - n^{\frac{3}{4}}- \log n} \label{lowerBoundonBProof} \\
& \stackrel{\eqref{sufficientLarge2}}{\ge} n \label{lowerBoundonB}
\end{align}
 and
\begin{align}
M & \ge \min\left\{e^{n \mathrm{C}\left(\frac{(1-n^{-1/4})\tilde \alpha P_1^{(n)}}{N_2}\right) - n^{\frac{3}{4}}-\log n},
 e^{ n\mathrm{C}\left(\frac{(1-n^{-1/4})\left(P_1^{(n)}+P_2^{(n)}+2\sqrt{(1-\tilde \alpha)P_1^{(n)}P_2^{(n)}}\right)}{ N_2+N_3}\right) - 2n^{3/4} - 3\log n}\right\}  \label{lowerBoundonMProof}\\
 &\stackrel{\eqref{defAlpha*}}{\ge} e^{n \mathrm{C}\left(\frac{(1-n^{-1/4})\tilde \alpha P_1^{(n)}}{N_2}\right) - 2n^{3/4} - 3\log n} \label{defM*} \\*
 &\stackrel{\eqref{sufficientLarge3}}{\ge}n^{1/4},
\label{lowerBoundonM}
\end{align}
where the detailed calculations of~\eqref{lowerBoundonBProof} and~\eqref{lowerBoundonMProof} are relegated to Appendix~\ref{appendixA}.
On the other hand, we have
\begin{align}
(M-1) e^{-\left(n\E\left[\log\left(\frac{p(Y_2|U,V)}{p(Y_2|V)}\right)\right] - \sqrt{n^{\frac{3}{2}}\Var\left[\log\left(\frac{p(Y_2|U,V)}{p(Y_2|V)}\right)\right]}\right)} \stackrel{\eqref{defM}}{\le} \frac{1}{n}, \label{errorProbCalEq8}
\end{align}
\begin{align}
(B-1) e^{-\left( n\E\left[\log\left(\frac{p(Y_3|V)}{p(Y_3)}\right)\right] - \sqrt{n^{\frac{3}{2}}\Var\left[\log\left(\frac{p(Y_3|V)}{p(Y_3)}\right)\right]}\right)} \stackrel{\eqref{defB}}{\le} \frac{1}{n}  \label{errorProbCalEq9}
\end{align}
and
\begin{align}
&\left(\frac{nM}{B}-1\right)e^{-\left( n\E\left[\log\left(\frac{p(Y_3|U,V)}{p(Y_3|V)}\right)\right] - \sqrt{n^{\frac{3}{2}}\Var\left[\log\left(\frac{p(Y_3|U,V)}{p(Y_3|V)}\right)\right]} \right)} \notag\\
& \stackrel{\eqref{lowerBoundonB}}{\le} \left(\frac{n(M-1)}{B}\right) e^{-\left( n\E\left[\log\left(\frac{p(Y_3|U,V)}{p(Y_3|V)}\right)\right] - \sqrt{n^{\frac{3}{2}}\Var\left[\log\left(\frac{p(Y_3|U,V)}{p(Y_3|V)}\right)\right]} \right)}\notag\\
& \stackrel{\eqref{defB}}{\le}  (M-1)\, e^{-\left( n\E\left[\log\left(\frac{p(Y_3|U,V)}{p(Y_3)}\right)\right] - \sqrt{n^{\frac{3}{2}}\Var\left[\log\left(\frac{p(Y_3|V)}{p(Y_3)}\right)\right]} - \sqrt{n^{\frac{3}{2}}\Var\left[\log\left(\frac{p(Y_3|U,V)}{p(Y_3|V)}\right)\right]} -2\log n\right)}\notag\\
& \stackrel{\eqref{defM}}{\le} \frac{1}{n}.  \label{errorProbCalEq10}
\end{align}
Combining the upper bound on the error probability of decoding the overall message in~\eqref{errorProbCalEq1}, the upper bound on the decoding error probabilities of submessages at the relay in \eqref{errorProbCalEq3Temp1*} and \eqref{errorProbCalEq3Temp1}, the upper bound on the error probability of decoding the overall message at the destination in  \eqref{errorProbCalEq5}, the upper bound on the decoding error probabilities of bin indices and submessages at the destination in~\eqref{errorProbCalEq6} and~\eqref{errorProbCalEq7} respectively, and the three upper bounds~\eqref{errorProbCalEq8}, \eqref{errorProbCalEq9} and~\eqref{errorProbCalEq10} obtained by the careful choices of~$B$ and~$M$, we obtain
\begin{align}
 \Pr\left\{\boldsymbol{W} \ne \hat{\boldsymbol{W}}\right\}
 \le L\left(\frac{13}{n^{1/2}}+\frac{4}{n}\right)  +\varepsilon - \frac{34}{n^{1/4}}.\label{errorProbCalEq11}
\end{align}
Setting
\begin{equation}
L\triangleq \lceil n^{1/4}\rceil, \label{defL}
\end{equation}
it follows from~\eqref{errorProbCalEq11} that
\begin{equation}
\Pr\left\{\boldsymbol{W} \ne \hat{\boldsymbol{W}}\right\} \le \varepsilon. \label{errorProbCalEq12}
\end{equation}
\subsection{Calculation of Average Power} \label{subsecPower}
In order to prove that the expected power constraint is satisfied for the source, we consider the following chain of inequalities:
\begin{align}
\E\left[\|X_1^{(L+1)n}\|^2\right]
 &\stackrel{\text{(a)}}{\le}nP_1^{(n)} + Ln P_1^{(n)}\left(\frac{(1-\varepsilon+34n^{-1/4})M+1}{M}\right)\notag\\
& = (L+1)n P_1^{(n)} \left( \frac{1+L\left(1-\varepsilon+34n^{-1/4} + M^{-1}\right)}{L+1} \right)\notag\\
& = (L+1)n P_1^{(n)} \left( \frac{\varepsilon-34n^{-1/4} - M^{-1}}{L+1}+1-\varepsilon+34n^{-1/4} + M^{-1} \right)\notag\\
&\stackrel{\eqref{lowerBoundonM}}{\le} (L+1)n P_1^{(n)} \left( \frac{\varepsilon-34n^{-1/4}}{L+1}+1-\varepsilon+35n^{-1/4} \right) \notag\\
&\stackrel{\text{(b)}}{\le} (L+1)n P_1^{(n)}(1-\varepsilon+36n^{-1/4}) \notag \\
& \stackrel{\eqref{defPin}}{\le}(L+1)n P_1\label{averagePowerEq1}
\end{align}
where
\begin{enumerate}
\item[(a)] follows from the definition of~$\mathcal{A}$ in~\eqref{defSetA} and the transmission rule used by the source in~\eqref{sourceTransmissionRule}.
\item[(b)] follows from the fact that $0<\varepsilon-34n^{-1/4}<1$ by~\eqref{sufficientLarge1} and the fact that $L\ge n^{1/4}$ by~\eqref{defL}.
\end{enumerate}
It remains to show that the expected power constraint is satisfied for the relay. Consider the following chain of inequalities:
\begin{align}
\E\left[\|X_2^{(L+1)n}\|^2\right]
& = \E\left[\left.\|X_2^{(L+1)n}\|^2\right|\mathcal{G}_2\right]\Pr\left\{\mathcal{G}_2\right\} + \E\left[\left.\|X_2^{(L+1)n}\|^2\right|\mathcal{G}_2^c\right]\Pr\left\{\mathcal{G}_2^c\right\} \notag\\
& \stackrel{\eqref{relayDecodeForward}}{\le} Ln P_2^{(n)}\Pr\left\{\mathcal{G}_2\right\}. \label{averagePowerEq2}
\end{align}
In order to bound $\Pr\left\{\mathcal{G}_2\right\}$, we write
\begin{align}
\Pr\left\{\mathcal{G}_2\right\}
&= \Pr\left\{\mathcal{G}_2\cap \mathcal{G}_1\right\} + \Pr\left\{\mathcal{G}_2\cap \mathcal{G}_1^c\right\} \notag \\
&\le \Pr\{\mathcal{G}_1\} + \Pr\left\{\mathcal{H}_{1}^c\right\} \notag\\
& \stackrel{\eqref{defSetA}}{\le} 1-\varepsilon+34n^{-1/4}+M^{-1} + \Pr\left\{\mathcal{H}_{1}^c\right\} \notag\\
&\stackrel{\eqref{lowerBoundonM}}{\le}1-\varepsilon + 35n^{-1/4} + \Pr\left\{\mathcal{H}_{1}^c\right\}. \label{averagePowerEq3}
\end{align}
Consider the following chain of inequalities:
\begin{align}
\Pr\left\{\mathcal{H}_{1}^c\right\}
& \stackrel{\text{(a)}}{\le} \Pr\left\{\mathcal{H}_{1}^c\cap \mathcal{E}_{1,1}\right\} + \Pr\left\{\mathcal{E}_{1,1}^c\right\} \notag\\
&\stackrel{\eqref{powerOutage}}{\le}  \Pr\left\{\mathcal{H}_{1}^c\cap \mathcal{E}_{1,1}\right\} + \frac{2}{n^{1/2}}\notag\\
&\stackrel{\text{(b)}}{\le}\Pr\left\{\left.\mathcal{H}_{1}^c\cap \mathcal{F}_{1,1}\right|\mathcal{H}_0\right\} + \frac{2}{n^{1/2}}\notag\\
& \stackrel{\text{(c)}}{\le} \frac{1}{n} + \frac{3}{n^{1/2}}. \label{averagePowerEq4}
\end{align}
where
\begin{enumerate}
\item[(a)] follows from the union bound.
\item[(b)] follows from the transmission rule~\eqref{sourceTransmissionRule} used by the source and the convention in~\eqref{conventionW*}.
    \item[(c)] follows from \eqref{errorProbCalEq3Temp1} and~\eqref{errorProbCalEq8}.
\end{enumerate}
  Combining~\eqref{averagePowerEq3} and \eqref{averagePowerEq4}, we have
\begin{align}
\Pr\left\{\mathcal{G}_2\right\} \le 1-\varepsilon + 39n^{-1/4}, \label{averagePowerEq5}
\end{align}
which implies from~\eqref{averagePowerEq2} and~\eqref{defPin} that
\begin{align}
\E\left[\|X_2^{(L+1)n}\|^2\right]  \le Ln P_2. \label{averagePowerEq6}
\end{align}
\subsection{Calculation of the $\varepsilon$-Achievable Rate}\label{subsecRate}
Combining~\eqref{errorProbCalEq12}, \eqref{averagePowerEq1} and~\eqref{averagePowerEq6}, we conclude that the code constructed above is an $((L+1)n, M^L, P_1, P_2, \varepsilon)$-code. In the following, we will obtain a lower bound on
$
\frac{L\log M}{(L+1)n} $, which is the rate of the $((L+1)n, M^L, P_1, P_2, \varepsilon)$-code, in terms of $\mathrm{C}\left(\frac{\tilde \alpha P_1}{(1-\varepsilon)N_2}\right)$.
To this end, we first use~\eqref{defM*} and the definitions of $P_1^{(n)}$ and $P_2^{(n)}$ in~\eqref{defPin} to obtain
\begin{align}
\log M \ge
n\mathrm{C}\left(\frac{(1-n^{-1/4})\tilde \alpha P_1}{ (1-\varepsilon+39n^{-1/4})N_2}\right) - 2n^{3/4} - 3\log n. \label{achRateEq1}
\end{align}
Since
\begin{align*}
(1-\varepsilon)\left(\frac{1-n^{-1/4}}{1-\varepsilon+39n^{-1/4}}\right) & = \left(1-n^{-1/4}\right)\left(1-\frac{39 n^{-1/4}}{1-\varepsilon+39n^{-1/4}}\right) \\
& \ge \left(1-n^{-1/4}\right)\left(1-\frac{39 n^{-1/4}}{1-\varepsilon}\right)\\
&\ge 1- \left(1+\frac{39}{1-\varepsilon}\right)n^{-1/4},
\end{align*}
it follows from~\eqref{defCapacityFunction}, \eqref{sufficientLarge1*} and the inequality
\begin{equation*}
\log(1+a-b)\ge \log(1+a) - \frac{b}{1+a-b} \ge \log(1+a) -b
 \end{equation*}
 for any $a>b>0$ based on Taylor's theorem that
\begin{align}
\mathrm{C}\left(\frac{(1-n^{-1/4})\tilde \alpha P_1}{(1-\varepsilon+39n^{-1/4})N_2}\right)
\ge\mathrm{C}\left( \frac{\tilde \alpha P_1}{(1-\varepsilon)N_2}\right) - \frac{1}{2}\left(1+\frac{39}{1-\varepsilon}\right)\left( \frac{\tilde \alpha P_1}{(1-\varepsilon)N_2}\right)n^{-1/4}. \label{achRateEq2}
\end{align}
Defining
\begin{equation}
\kappa_1\triangleq \frac{1}{2}\left(1+\frac{39}{1-\varepsilon}\right)\left( \frac{\tilde \alpha P_1}{(1-\varepsilon)N_2}\right) + 5 \label{defKappa1}
\end{equation}
and combining~\eqref{achRateEq1} and~\eqref{achRateEq2}, we have
\begin{align}
\log M
&\ge n\mathrm{C}\left(\frac{\tilde \alpha P_1}{(1-\varepsilon)N_2}\right) - \kappa_1 n^{3/4}. \label{achRateEq3}
\end{align}
In order to bound the rate of the constructed code $\frac{\log M}{(L+1)n}$, we consider
\begin{align}
L \log M &= (L+1)\left(\log M - \frac{\log M}{L+1}\right)\notag\\
&\stackrel{\eqref{defL}}{\ge} (L+1)\left(\log M - \frac{\log M}{n^{1/4}}\right). \label{achRateEq4}
\end{align}
 Recalling~\eqref{eqnCapacityC}, we follow the standard converse proof argument based on Fano's inequality \cite[Sec.~16.2]{elgamal} and obtain
\begin{align}
\log M \le \frac{nC + 1}{1-\varepsilon}. \label{achRateEq4*}
\end{align}
 Defining
\begin{equation}
\kappa_2 \triangleq \frac{C + 1}{1-\varepsilon} \label{defKappa2}
\end{equation}
and combining~\eqref{achRateEq4} and~\eqref{achRateEq4*}, we have
\begin{align}
L \log M &\ge (L+1)\left(\log M - \kappa_2 n^{3/4}\right). \label{achRateEq5}
\end{align}
In addition,
\begin{align}
(L+1)n^{3/4} & \stackrel{\eqref{defL}}{\le} n+ 2n^{3/4}\notag\\
&\le 3n \notag\\
&\stackrel{\eqref{defL}}{\le} 3\left((L+1)n\right)^{4/5}. \label{achRateEq6}
\end{align}
Consequently, it follows from~\eqref{achRateEq5}, \eqref{achRateEq3}, \eqref{achRateEq6} and the definition of~$L$ in~\eqref{defL} that
\begin{align}
\lceil n^{1/4}\rceil \log M &\ge (\lceil n^{1/4}\rceil+1)n \mathrm{C}\left(\frac{\tilde \alpha P_1}{(1-\varepsilon)N_2}\right)  - 3(\kappa_1+\kappa_2)\left((\lceil n^{1/4}\rceil+1)n\right)^{4/5}. \label{achRateEq7}
\end{align}
Although the numbers of channel uses $(\lceil n^{1/4}\rceil+1)n$ are not consecutive integers as $n$ increases, we can construct a sequence of $(m, M_m, P_1, P_2, \varepsilon)$-codes based on the $((L+1)n, M^L, P_1, P_2, \varepsilon)$-codes such that for all sufficiently large~$m\in \mathbb{N}$,
\begin{align}
\log M_m \ge  m\max\limits_{0\le\alpha\le 1}R_{\text{cut-set}}\left(\alpha, \frac{P_1}{1-\varepsilon}, \frac{P_2}{1-\varepsilon}\right) - \kappa_3 m^{4/5} \label{achRateEq8}
\end{align}
for some $\kappa_3>0$ which is a function of $(\varepsilon, P_1, P_2, N_1, N_2)$ and does not depend on~$m$. To this end, we define for each $m\in\mathbb{N}$ an $(m, M_m, P_1, P_2, \varepsilon)$-code to be identical to the constructed  $((L+1)n, M^L, P_1, P_2, \varepsilon)$-code with $(L+1)n$ chosen to be as close to~$m$ as possible but not larger than~$m$. More specifically, we define for each $m\in\mathbb{N}$ an $n_m$ to be the unique natural number that satisfies
 \begin{equation}
(\lceil n_m^{1/4}\rceil +1)n_m \le m < \lceil (n_m+1)^{1/4}\rceil+1)(n_m+1), \label{defNM}
\end{equation}
define
\begin{equation}
M_m\triangleq M^{\lceil n_m^{1/4}\rceil}, \label{defMm}
\end{equation}
and define for each $m\in\mathbb{N}$ an $(m, M_m, P_1, P_2, \varepsilon)$-code which is identical to the constructed \linebreak
$((\lceil n_m^{1/4}\rceil+1)n_m, M^{\lceil n_m^{1/4}\rceil}, P_1, P_2, \varepsilon)$-code.
Then for each sufficiently large~$m$,
\begin{align}
\log M_m &\stackrel{\eqref{defMm}}{=}\lceil n_m^{1/4} \rceil\log M \notag\\
&\stackrel{\eqref{achRateEq7}}{\ge}(\lceil n_m^{1/4}\rceil+1)n_m \mathrm{C}\left(\frac{\tilde \alpha P_1}{(1-\varepsilon)N_2}\right)  - 3(\kappa_1+\kappa_2)\left((\lceil n_m^{1/4}\rceil+1)n_m\right)^{4/5}\notag\\
&\stackrel{\text{(a)}}{\ge} (\lceil (n_m+1)^{1/4}\rceil+1)(n_m+1) \mathrm{C}\left(\frac{\tilde \alpha P_1}{(1-\varepsilon)N_2}\right)  - 3(\kappa_1+\kappa_2)\left((\lceil n_m^{1/4}\rceil+1)n_m\right)^{4/5}\notag\\
&\qquad -  (1+2n_m+\lceil (n_m+1)^{1/4}\rceil)\mathrm{C}\left(\frac{\tilde \alpha P_1}{(1-\varepsilon)N_2}\right) \notag\\
&\stackrel{\eqref{defNM}}{\ge} m \mathrm{C}\left(\frac{\tilde \alpha P_1}{(1-\varepsilon)N_2}\right)  - 3(\kappa_1+\kappa_2)m^{4/5}-  (1+2n_m+\lceil (n_m+1)^{1/4}\rceil)\mathrm{C}\left(\frac{\tilde \alpha P_1}{(1-\varepsilon)N_2}\right)
\notag\\
&\stackrel{\text{(b)}}{\ge} m \mathrm{C}\left(\frac{\tilde \alpha P_1}{(1-\varepsilon)N_2}\right)  - 3(\kappa_1+\kappa_2)m^{4/5}-  (1+2m^{4/5}+\lceil (m^{4/5}+1)^{1/4}\rceil)\mathrm{C}\left(\frac{\tilde \alpha P_1}{(1-\varepsilon)N_2}\right)\label{achRateEq9}
\end{align}
where
\begin{enumerate}
\item[(a)] follows from the fact that
\begin{align*}
(\lceil n_m^{1/4}\rceil+1)n_m &\ge ( n_m^{1/4}+1)n_m \\
& \ge (n_m+1)^{1/4}n_m \\
& \ge \left(\left\lceil (n_m+1)^{1/4}\right\rceil -1\right)n_m \\
& =  \left(\left\lceil(n_m+1)^{1/4}\right\rceil +1\right)(n_m+1) -\left\lceil(n_m+1)^{1/4}\right\rceil -2n_m -1.
\end{align*}
\item[(b)] follows from \eqref{defNM} that $n_m^{5/4}\le m$.
\end{enumerate}
Recalling the choices of $\kappa_1$ and $\kappa_2$ in~\eqref{defKappa1} and~\eqref{defKappa2} respectively, it follows from~\eqref{achRateEq9} and~\eqref{alpha*St} that there exists a $\kappa_3>0$ that is a function of $(\varepsilon, P_1, P_2, N_1, N_2)$ and does not depend on~$m$ such that~\eqref{achRateEq8} holds for all sufficiently large~$m$. Consequently,~\eqref{achRateEq8} holds for each $\varepsilon\in(0,1)$ for all sufficiently large~$m$, which implies~\eqref{achProofSt}.

\section{Converse Proof of Theorem~\ref{thmMainResult}}\label{secConverse}
In this section, we will show that for all $\varepsilon\in(0,1)$
\begin{equation}
C_\varepsilon \le  \max\limits_{0\le\alpha\le 1}R_{\text{cut-set}}\left(\alpha, \frac{P_1}{1-\varepsilon}, \frac{P_2}{1-\varepsilon}\right). \label{convProofSt}
\end{equation}
To this end, we fix an $\varepsilon\in (0,1)$ and let $R$ be an $\varepsilon$-achievable rate. By Definitions~\ref{defAchievableRate} and~\ref{defCapacity}, there exists a sequence of $(n, M_n, P_1, P_2, \varepsilon)$-codes such that
\begin{equation}
\liminf_{n\rightarrow\infty}\frac{1}{n}\log M_n \ge R. \label{thmTempEq2}
\end{equation}
Fix any sufficiently large~$n\in \mathbb{N}$ such that
 \begin{align}
\frac{1}{\sqrt{n}}\le \frac{1-\varepsilon}{2} \label{sufficientLarge1Converse}
 \end{align}
 and
 \begin{align}
 \frac{1}{n}\,e^{\left(\frac{4}{1-\varepsilon}\right)\left(\frac{P_1+P_2+\sqrt{P_1P_2}}{N_2}\right)\left(\frac{2(P_1+P_2+\sqrt{P_1P_2})}{(1-\varepsilon)N_2}+1\right)} < \frac{1}{2\sqrt{n}}, \label{sufficientLarge2Converse}
 \end{align}
 and let $p_{W,X_1^n, X_2^n, Y_2^n, Y_3^n, \hat W}$ be the probability distribution induced by the $(n, M_n, P_1, P_2, \varepsilon)$-code. To simplify notation, we omit the subscripts of the probability and expectation terms which are evaluated according to~$p_{W,X_1^n, X_2^n, Y_2^n, Y_3^n, \hat W}$ in the rest of Section~\ref{secConverse}.
\subsection{Obtaining a Lower Bound on the Error Probability in Terms of the Type-II Errors of Binary Hypothesis Tests}
Define
\begin{align}
 \rho \triangleq \frac{\sum_{k=1}^n \E[X_{1,k}X_{2,k}]}{n \sqrt{P_1 P_2}} \label{defRho}
 \end{align}
 where
 \begin{equation}
 \rho^2 \le 1 \label{defRho*}
 \end{equation}
 by Cauchy-Schwartz inequality and the power constraint~\eqref{powerConstraint} for each $i\in\{1,2\}$,
 and define
 \begin{align}
 P_i^{(n)}&\triangleq \frac{P_i}{1-\varepsilon-n^{-1/2}} \label{defPinConverse}\\
 & \stackrel{\eqref{sufficientLarge1Converse}}{\le} \frac{2P_i}{1-\varepsilon} \label{defPinConverse*}
 \end{align}
 for each $i\in\{1,2\}$. In addition, define
\begin{align}
s_{Y_3^{n}, \hat W}^{(1)}\triangleq \left(\prod_{k=1}^{n} s_{Y_{3,k}}^{(1)} \right)p_{\hat W|Y_3^{n}} \label{defDistTildeS}
\end{align}
where
\begin{align}
s_{Y_{3,k}}^{(1)}(y_{3,k})\triangleq \mathcal{N}\left(y_{3,k}; 0, P_1^{(n)}+P_2^{(n)}+2\rho\sqrt{P_1^{(n)}P_2^{(n)}}+N_2+N_3\right), \label{defTildeSy3}
\end{align}
and define
\begin{align}
s_{X_2^{n}, Y_2^{n}, Y_3^{n}, \hat W}^{(2)}\triangleq \left(\prod_{k=1}^{n}  p_{X_{2,k}|Y_2^{k-1}, Y_3^{k-1}} s_{Y_{2,k}|X_{2,k}}^{(2)} p_{Y_{3,k}|X_{2,k},Y_{2,k}} \right) p_{\hat W|Y_3^{n}} \label{defDistBarS}
\end{align}
where
\begin{align}
s_{Y_{2,k}|X_{2,k}}^{(2)}(y_{2,k}|x_{2,k})\triangleq \mathcal{N}\left(y_{2,k}; \left(\frac{\E[X_{1,k}X_{2,k}]}{\E[X_{2,k}^2]}\right)x_{2,k}, (1-\rho^2) P_1^{(n)} +N_2 \right) \label{defBarSy2Givenx2}
\end{align}
with the convention
\begin{align}
\left(\frac{\E[X_{1,k}X_{2,k}]}{\E[X_{2,k}^2]}\right)x_{2,k} \triangleq 0 \qquad \text{if $\E_{p_{X_{2,k}}}[X_{2,k}^2]=0$}. \label{convention1Converse}
\end{align}
It follows from Proposition~\ref{propositionBHTLowerBound} and Definition~\ref{defCode}
with the identifications $U\equiv W$, $V\equiv \hat W$, $p_{U,V}\equiv  p_{W,  \hat W}$, $|\mathcal{W}|\equiv  M_n$ and
\begin{equation}
\alpha\equiv \Pr\{W\ne  \hat W\} \le \varepsilon \label{alphaLessThanEpsilon}
\end{equation}
 that for each $j\in\{1,2\}$,
 \begin{align}
\beta_{1- \varepsilon}(p_{W,\hat W}\|p_{W} s_{\hat W}^{(j)})
 & \stackrel{\eqref{alphaLessThanEpsilon}}{\le} \beta_{1-\alpha}(p_{W,\hat W}\|p_{W} s_{\hat W}^{(j)}) \notag\\
  & \le \frac{1}{M_n}.\label{eqnBHTReverseChain}
 \end{align}

    \subsection{Using the DPI to Introduce the Channel Inputs and Outputs}
Using the DPI of $\beta_{1- \varepsilon}$ in Lemma~\ref{lemmaDPI}, we have
 \begin{align}
& \beta_{1- \varepsilon}(p_{W,\hat W}\|p_{W} s_{\hat W}^{(1)})\notag\\
& \ge \beta_{1- \varepsilon}\left(p_{W,X_1^n, X_2^n, Y_3^n, \hat W}\left\|p_{W}\left(\prod_{k=1}^n p_{X_{1,k}, X_{2,k}|W, X_1^{k-1}, X_2^{k-1}, Y_3^{k-1}}\right) s_{Y_3^n,\hat W}^{(1)}\right.\right). \label{converseProofEq1}
\end{align}
Fix a $\xi_n^{(1)}>0$ to be specified later. Since
\begin{align*}
p_{W,X_1^n, X_2^n, Y_3^n, \hat W} &\stackrel{\eqref{memorylessStatement}}{=} p_{W} \left(\prod_{k=1}^n p_{X_{1,k}, X_{2,k}, Y_{3,k}|W, X_1^{k-1}, X_2^{k-1}, Y_3^{k-1}}\right)p_{\hat W|Y_3^n}\\
&\stackrel{\eqref{memorylessStatement*}}{=} p_W\left(\prod_{k=1}^n p_{X_{1,k}, X_{2,k}|W, X_1^{k-1}, X_2^{k-1}, Y_3^{k-1}}p_{Y_{3,k}|X_{1,k}, X_{2,k}}\right)p_{\hat W|Y_3^n},
 \end{align*}
it follows from \eqref{eqnBHTReverseChain}, the definition of $s_{Y_3^n,\hat W}^{(1)}$ in~\eqref{defDistTildeS}, \eqref{converseProofEq1} and Lemma~\ref{lemmaDPI} that
\begin{align}
\log M_n \le \log \xi_n^{(1)} - \log\left(1-\varepsilon - \Pr\left\{\sum_{k=1}^n \log\left( \frac{p_{Y_{3,k}|X_{1,k}, X_{2,k}}(Y_{3,k}|X_{1,k}, X_{2,k})}{s_{Y_{3,k}}^{(1)}(Y_{3,k})}\right)\ge \log\xi_n^{(1)}\right\}\right). \label{converseProofEq2}
\end{align}
On the other hand, it follows from the DPI of $\beta_{1- \varepsilon}$ in Lemma~\ref{lemmaDPI} that
 \begin{align}
\beta_{1- \varepsilon}(p_{W,\hat W}\|p_{W} s_{\hat W}^{(2)})\ge \beta_{1- \varepsilon}\left(p_{W,X_1^n, X_2^n, Y_2^n, Y_3^n, \hat W}\left\|p_{W}\left(\prod_{k=1}^np_{X_{1,k}|W, Y_2^{k-1}, Y_3^{k-1}} \right) s_{X_2^n, Y_2^n, Y_3^n,\hat W}^{(2)}\right.\right). \label{converseProofEq3}
\end{align}
Fix a $\xi_n^{(2)}>0$ to be specified later. Combining \eqref{converseProofEq3}, the factorization of $p_{W,X_1^n, X_2^n, Y_2^n, Y_3^n, \hat W}$ in~\eqref{memorylessStatement**}, the definition of $s_{X_2^n, Y_2^n, Y_3^n,\hat W}^{(2)}$ in~\eqref{defDistBarS}, \eqref{eqnBHTReverseChain} and Lemma~\ref{lemmaDPI}, we obtain
\begin{align}
\log M_n \le \log \xi_n^{(2)} - \log\left(1-\varepsilon - \Pr\left\{\sum_{k=1}^n \log\left( \frac{p_{Y_{2,k}|X_{1,k}}(Y_{2,k}|X_{1,k})}{s_{Y_{2,k}|X_{2,k}}^{(2)}(Y_{2,k}|X_{2,k})}\right)\ge \log\xi_n^{(2)}\right\}\right). \label{converseProofEq4}
\end{align}
\subsection{Simplifying the Information Spectrum Terms}
Let
\begin{align}
Z_2^n \triangleq Y_2^n - X_1^n \sim \mathcal{N}(z_2^n; 0, N_2)\label{defZ2n}
\end{align}
and
\begin{align}
Z_3^n \triangleq Y_3^n - Y_2^n - X_2^n \sim \mathcal{N}(z_3^n; 0, N_3)  \label{defZ3n}
\end{align}
be the Gaussian noises added at nodes~2 and~3 respectively (cf.\ \eqref{introY2k} and~\eqref{introY3k}), and define
\begin{align}
&p_{X_1^n, X_2^n, Y_2^n, Y_3^n, Z_2^n, Z_3^n}(x_1^n, x_2^n, y_2^n, y_3^n, z_2^n, z_3^n) \notag\\
&\quad \triangleq p_{X_1^n, X_2^n, Y_2^n, Y_3^n}(x_1^n, x_2^n, y_2^n, y_3^n)\mathbf{1}\left\{z_2^n = y_2^n - x_1^n\right\}\mathbf{1}\left\{z_3^n = y_3^n - y_2^n - x_2^n\right\}. \label{memorylessStatementConverseProof}
\end{align}
Straightforward calculations based on Definition~\ref{defchannel} and~\eqref{memorylessStatementConverseProof} reveal that
\begin{align}
p_{X_1^n, X_2^n, Y_3^n} = \prod_{k=1}^n p_{X_{1,k}, X_{2,k}|X_1^{k-1}, X_2^{k-1}, Y_3^{k-1}}p_{Y_{3,k}|X_{1,k}, X_{2,k}}, \label{defDistPx1x2y3}
\end{align}
\begin{align}
p_{X_1^n, X_2^n, Z_2^n, Z_3^n} = \prod_{k=1}^n p_{X_{1,k}, X_{2,k}|X_1^{k-1}, X_2^{k-1}, Z_2^{k-1}, Z_3^{k-1}}p_{Z_{2,k}}p_{Z_{3,k}}, \label{defDistPx1x2z2z3}
\end{align}
and
\begin{align}
p_{X_1^n, X_2^n, Y_2^n} = \prod_{k=1}^n p_{X_{1,k}, X_{2,k}|X_1^{k-1}, X_2^{k-1}, Y_2^{k-1}}p_{Y_{2,k}|X_{1,k}}, \label{defDistPx1x2y2}
\end{align}
where
\begin{align}
p_{Y_{3,k}|X_{1,k}, X_{2,k}}(y_{3,k}|x_{1,k}, x_{2,k}) = \mathcal{N}(y_{3,k}-x_{1,k}-x_{2,k}; 0, N_2 + N_3) \label{defDistPy3GivenX1X2}
\end{align}
and
\begin{align}
p_{Y_{2,k}|X_{1,k}}(y_{2,k}|x_{1,k}) = \mathcal{N}(y_{2,k}-x_{1,k}; 0, N_2) \label{defDistPy2GivenX1X2}
\end{align}
for each $k\in\{1, 2, \ldots, n\}$.
In order to further simplify the expressions in $\eqref{converseProofEq2}$ and $\eqref{converseProofEq4}$, we define
\begin{align}
i_k^{(1)}&\triangleq \log\left( \frac{p_{Y_{3,k}|X_{1,k}, X_{2,k}}(Y_{3,k}|X_{1,k}, X_{2,k})}{s_{Y_{3,k}}^{(1)}(Y_{3,k})}\right),\label{defIk1}\\
i_k^{(2)}&\triangleq \log\left( \frac{p_{Y_{2,k}|X_{1,k}}(Y_{2,k}|X_{1,k})}{s_{Y_{2,k}|X_{2,k}}^{(2)}(Y_{2,k}|X_{2,k})}\right),
\label{defIk2}\\
\tilde X_k^{(1)}&\triangleq \frac{X_{1,k}+X_{2,k}}{\sqrt{N_2+N_3}}, \label{defTildeX1Converse} \\
\tilde X_k^{(2)}&\triangleq \frac{X_{1,k}-\left(\frac{\E[X_{1,k}X_{2,k}]}{\E[X_{2,k}^2]}\right)X_{2,k}}{\sqrt{N_2}} \qquad \text{(cf.\ \eqref{convention1Converse})},\label{defTildeX2Converse} \\
\tilde Z_k^{(1)}&\triangleq \frac{Z_{2,k}+Z_{3,k}}{\sqrt{N_2+N_3}},\label{defTildeZk1}\\
\tilde Z_k^{(2)}&\triangleq \frac{Z_{2,k}}{\sqrt{N_2}},\label{defTildeZk2}\\
\tilde P^{(1,n)}&\triangleq \frac{P_1^{(n)}+P_2^{(n)}+2\rho\sqrt{P_1^{(n)}P_2^{(n)}}}{N_2+N_3}\label{defTildeP1n}\\
&\text{and} \notag\\
\tilde P^{(2,n)}&\triangleq \frac{(1-\rho^2)P_1^{(n)}}{N_2}.\label{defTildeP2n}
\end{align}
Then, recalling \eqref{defTildeSy3}, \eqref{defBarSy2Givenx2}, \eqref{defDistPx1x2y3}, \eqref{defDistPx1x2z2z3}, \eqref{defDistPx1x2y2}, \eqref{defDistPy3GivenX1X2} and~\eqref{defDistPy2GivenX1X2},
we can rewrite the probabilities in~\eqref{converseProofEq2} and~\eqref{converseProofEq4} via the substitutions
\begin{equation*}
z_2^n + z_3^n = y_3^n-x_1^n + x_2^n
\end{equation*}
and
\begin{equation*}
z_2^n = y_2^n-x_1^n
\end{equation*}
 as
\begin{align}
\Pr\left\{\sum_{k=1}^n i_k^{(i)}\ge \log \xi_n^{(i)}\right\}=\Pr\left\{n\mathrm{C}\left(\tilde P^{(i,n)}\right)+\sum_{k=1}^n \frac{-\tilde P^{(i,n)}\left(\tilde Z_k^{(i)}\right)^2 + 2\tilde X_k^{(i)}\tilde Z_k^{(i)}+\left(\tilde X_k^{(i)}\right)^2}{2\left(\tilde P^{(i,n)}+1\right)} \ge \log \xi_n^{(i)}\right\} \label{converseProofEq7}
\end{align}
 for each $i\in\{1,2\}$ respectively (recall the definition of $\mathrm{C}(\cdot)$ in~\eqref{defCapacityFunction}).
\subsection{Introducing Events Related to the Power of Linear Combinations of $X_1^n$ and $X_2^n$}
Following~\eqref{converseProofEq7},
we consider
\begin{align}
\mu_1&\triangleq \frac{1}{n}\E\left[\sum_{k=1}^n\left(\tilde X_k^{(1)}\right)^2\right]\label{defMu1*}\\
&\stackrel{\eqref{defTildeX1Converse}}{=}\frac{\frac{1}{n}\sum_{k=1}^n \left(\E[X_{1,k}^2] + 2\E[X_{1,k}X_{2,k}] + \E[X_{2,k}^2]\right)}{N_2+N_3}\notag\\
&\stackrel{\eqref{powerConstraint}}{\le}\frac{P_1 + P_2 + \frac{2}{n}\sum_{k=1}^n \E[X_{1,k}X_{2,k}]}{N_2+N_3} \notag\\
&\stackrel{\eqref{defRho}}{=}\frac{P_1+P_2+2\rho\sqrt{P_1P_2}}{N_2+N_3}\label{defMu1}
\end{align}
and
\begin{align}
\mu_2&\triangleq \frac{1}{n}\E\left[\sum_{k=1}^n\left(\tilde X_k^{(2)}\right)^2\right] \label{defMu2*}\\
&\stackrel{\text{(a)}}{=}\frac{\frac{1}{n}\sum_{k=1}^n \left(\E[X_{1,k}^2] - \frac{(\E[X_{1,k}X_{2,k}])^2}{\E[X_{2,k}^2]}\right)}{N_2}\notag\\
&\stackrel{\eqref{powerConstraint}}{\le} \frac{P_1 - \frac{1}{n}\sum_{k=1}^n \frac{(\E[X_{1,k}X_{2,k}])^2}{\E[X_{2,k}^2]}}{N_2}\notag\\
& \stackrel{\text{(b)}}{\le} \frac{P_1(1-\rho^2)}{N_2}, \label{defMu2}
\end{align}
where
\begin{enumerate}
\item[(a)] follows from~\eqref{defTildeX2Converse} and the convention in~\eqref{convention1Converse}.
\item[(b)] is due to the following fact:
\begin{align*}
n\rho^2 P_1 &\stackrel{\eqref{defRho}}{=} \frac{\left(\sum_{k=1}^n \E[X_{1,k}X_{2,k}]\right)^2}{nP_2}\\
& = \frac{\left(\sum_{k=1}^n \E[X_{1,k}X_{2,k}]\right)^2}{nP_2}\\
& = \frac{\left(\sum\limits_{k=1}^n \E\left[\left(\frac{X_{1,k}X_{2,k}}{\sqrt{\E[X_{2,k}^2]}}\right)\sqrt{\E[X_{2,k}^2]}\right]\right)^2}{nP_2}\\
& \le  \sum_{k=1}^n \frac{(\E[X_{1,k}X_{2,k}])^2}{\E[X_{2,k}^2]}
\end{align*}
where the inequality follows from Cauchy-Schwartz inequality and the power constraint in~\eqref{powerConstraint} for each $i\in\{1,2\}$.
\end{enumerate}
Let
\begin{align}
\mu_i^{(n)}\triangleq \frac{\mu_i}{1-\varepsilon-n^{-1/2}},\label{defMuIn}
\end{align}
and define
\begin{align}
\mathcal{E}_i\triangleq\left\{\sum_{k=1}^n\left(\tilde X_k^{(i)}\right)^2 \le n\mu_i^{(n)}\right\} \label{defEventEIConverse}
\end{align}
to be the event that the power of the specific linear combination of~$X_1^n$ and~$X_2^n$ does not exceed $\mu_i^{(n)}$.
Using equations~\eqref{defMu1*}, \eqref{defMu2*}, \eqref{defMuIn}, \eqref{defEventEIConverse} and Markov's inequality, we obtain that
\begin{align}
\Pr\{\mathcal{E}_i^c\}\le 1-\varepsilon - n^{-1/2} \label{converseProofEq8}
\end{align}
for each~$i\in\{1,2\}$. In order to bound the RHS of~\eqref{converseProofEq7}, we use the union bound and~\eqref{converseProofEq8} to obtain
\begin{align}
&\Pr\left\{n\mathrm{C}\left(\tilde P^{(i,n)}\right)+\sum_{k=1}^n \frac{-\tilde P^{(i,n)}\left(\tilde Z_k^{(i)}\right)^2 + 2\tilde X_k^{(i)}\tilde Z_k^{(i)}+\left(\tilde X_k^{(i)}\right)^2}{2\left(\tilde P^{(i,n)}+1\right)} \ge \log \xi_n^{(i)}\right\}  \notag\\
& \le \Pr\left\{\left\{n\mathrm{C}\left(\tilde P^{(i,n)}\right)+\sum_{k=1}^n \frac{-\tilde P^{(i,n)}\left(\tilde Z_k^{(i)}\right)^2 + 2\tilde X_k^{(i)}\tilde Z_k^{(i)}+\left(\tilde X_k^{(i)}\right)^2}{2\left(\tilde P^{(i,n)}+1\right)} \ge \log \xi_n^{(i)}\right\}\cap \mathcal{E}_i\right\} + 1-\varepsilon - n^{-1/2}. \label{converseProofEq9}
\end{align}
\subsection{Simplifying the Two Probability Terms Conditioned on Events $\mathcal{E}_1$ and $\mathcal{E}_2$ Respectively}
Following~\eqref{converseProofEq9} and letting
\begin{equation}
\gamma_n^{(i)}\triangleq  2\left(\tilde P^{(i,n)}+1\right)\left(\log\xi_n^{(i)} - n\mathrm{C}\left(\tilde P^{(i,n)}\right)\right) \label{defGammaI}
\end{equation}
for each $i\in\{1,2\}$,
we consider
\begin{align}
&\Pr\left\{\left\{n\mathrm{C}\left(\tilde P^{(i,n)}\right)+\sum_{k=1}^n \frac{-\tilde P^{(i,n)}\left(\tilde Z_k^{(i)}\right)^2 + 2\tilde X_k^{(i)}\tilde Z_k^{(i)}+\left(\tilde X_k^{(i)}\right)^2}{2\left(\tilde P^{(i,n)}+1\right)} \ge \log \xi_n^{(i)}\right\}\cap \mathcal{E}_i\right\}\notag\\
&\stackrel{\eqref{defGammaI}}{=}\Pr\left\{\left\{\sum_{k=1}^n \left(-\tilde P^{(i,n)}\left(\tilde Z_k^{(i)}\right)^2 + 2\tilde X_k^{(i)}\tilde Z_k^{(i)}+\left(\tilde X_k^{(i)}\right)^2\right) \ge \gamma_n^{(i)}\right\}\cap \mathcal{E}_i\right\}\notag\\
&\stackrel{\eqref{defEventEIConverse}}{\le} \Pr\left\{\left\{\sum_{k=1}^n \left(-\tilde P^{(i,n)}\left(\tilde Z_k^{(i)}\right)^2 + 2\tilde X_k^{(i)}\tilde Z_k^{(i)}+ \mu_i^{(n)} \right) \ge \gamma_n^{(i)}\right\}\cap \mathcal{E}_i\right\}\notag\\
&\stackrel{\text{(a)}}{\le}  \Pr\left\{\left\{\sum_{k=1}^n \left(-\tilde P^{(i,n)}\left(\tilde Z_k^{(i)}\right)^2 + 2\tilde X_k^{(i)}\tilde Z_k^{(i)}+\tilde P^{(i,n)}\right) \ge \gamma_n^{(i)}\right\}\cap \mathcal{E}_i\right\} \label{converseProofEq10}
\end{align}
where
\begin{enumerate}
\item[(a)] follows from the following inequality for each $i\in\{1,2\}$, which results from combining the definition of $\mu_i^{(n)}$ in~\eqref{defMuIn}, the bounds on $\mu_1$ and $\mu_2$ in \eqref{defMu1} and \eqref{defMu2} respectively, the definitions of $\tilde P^{(1,n)}$ and~$\tilde P^{(2,n)}$ in~\eqref{defTildeP1n} and~\eqref{defTildeP2n} respectively, and the definition of $P_i^{(n)}$ in~\eqref{defPinConverse}:
    \begin{equation}
   \mu_i^{(n)}\le \tilde P^{(i,n)}. \label{muILessThanPin}
    \end{equation}
\end{enumerate}
Define
\begin{equation}
t_n\triangleq n^{-1/2}. \label{defTn}
 \end{equation}
 Following~\eqref{converseProofEq10}, we consider the following chain of inequalities for each $i\in\{1,2\}$:
 \begin{align}
 & \Pr\left\{\left\{\sum_{k=1}^n \left(-\tilde P^{(i,n)}\left(\tilde Z_k^{(i)}\right)^2 + 2\tilde X_k^{(i)}\tilde Z_k^{(i)}+\tilde P^{(i,n)}\right) \ge \gamma_n^{(i)}\right\}\cap \mathcal{E}_i\right\} \notag\\
 & =  \Pr\left\{\left. t_n \sum_{k=1}^n \left(-\tilde P^{(i,n)}\left(\tilde Z_k^{(i)}\right)^2 + 2\tilde X_k^{(i)}\tilde Z_k^{(i)}+\tilde P^{(i,n)}\right) \ge t_n\gamma_n^{(i)}\right| \mathcal{E}_i\right\}\Pr\{\mathcal{E}_i\} \notag\\
 &\stackrel{\text{(a)}}{\le} \E\left[\left. e^{t_n\sum\limits_{k=1}^n\left(-\tilde P^{(i,n)}\left(\tilde Z_k^{(i)}\right)^2 + 2\tilde X_k^{(i)}\tilde Z_k^{(i)}+\tilde P^{(i,n)}\right)}\right|\mathcal{E}_i\right]\Pr\{\mathcal{E}_i\}e^{-t_n\gamma_n^{(i)}} \notag\\
 &\stackrel{\eqref{defEventEIConverse}}{\le} \E\left[\left. e^{t_n\sum\limits_{k=1}^n\left(-\tilde P^{(i,n)}\left(\tilde Z_k^{(i)}\right)^2 + 2\tilde X_k^{(i)}\tilde Z_k^{(i)}+\tilde P^{(i,n)}\right)+\frac{2t_n^2}{1+2t_n \tilde P^{(i,n)}}\left(n\mu_i^{(n)} - \sum\limits_{k=1}^n \left(\tilde X_k^{(i)}\right)^2 \right)}\right|\mathcal{E}_i\right]\Pr\{\mathcal{E}_i\}e^{-t_n\gamma_n^{(i)}} \notag\\
 &\stackrel{\eqref{muILessThanPin}}{\le}\E\left[ e^{t_n\sum\limits_{k=1}^n\left(-\tilde P^{(i,n)}\left(\tilde Z_k^{(i)}\right)^2 + 2\tilde X_k^{(i)}\tilde Z_k^{(i)}+\tilde P^{(i,n)}\right)+\frac{2t_n^2}{1+2t_n \tilde P^{(i,n)}}\left(n\tilde P^{(i,n)} - \sum\limits_{k=1}^n \left(\tilde X_k^{(i)}\right)^2 \right)}\right]e^{-t_n\gamma_n^{(i)}} \label{converseProofEq11}
 \end{align}
 where (a) follows from Markov's inequality. Combining~\eqref{defDistPx1x2z2z3}, the definitions of $\tilde X_k^{(1)}$ and $\tilde X_k^{(2)}$ in \eqref{defTildeX1Converse} and \eqref{defTildeX2Converse} respectively, the definitions of $Z_2^n$ and $Z_3^n$ in~\eqref{defZ2n} and~\eqref{defZ3n} respectively, and the definitions of $\tilde Z_k^{(1)}$ and $\tilde Z_k^{(2)}$ in~\eqref{defTildeZk1} and~\eqref{defTildeZk2} respectively, we conclude that
 that $\tilde Z_k^{(i)}\sim \mathcal{N}(\tilde z_k^{(i)}; 0, 1)$ and $\tilde Z_k^{(i)}$ and $\left(\{\tilde X_m^{(i)}\}_{m=1}^k, \{\tilde Z_\ell^{(i)}\}_{\ell=1}^{k-1}\right)$ are independent for each $i\in\{1,2\}$ and each $k\in\{1, 2, \ldots, n\}$, which implies that
 \begin{align}
 &\E\left[ e^{t_n\sum\limits_{k=1}^n\left(-\tilde P^{(i,n)}\left(\tilde Z_k^{(i)}\right)^2 + 2\tilde X_k^{(i)}\tilde Z_k^{(i)}+\tilde P^{(i,n)}\right)+\frac{2t_n^2}{1+2t_n \tilde P^{(i,n)}}\left(n\tilde P^{(i,n)} - \sum\limits_{k=1}^n \left(\tilde X_k^{(i)}\right)^2 \right)}\right]\notag\\
 & = (1+2t_n \tilde P^{(i,n)})^{\frac{-n}{2}}e^{nt_n\tilde P^{(i,n)} + \frac{2nt_n^2\tilde P^{(i,n)}}{1+2t_n \tilde P^{(i,n)}}},  \label{converseProofEq12}
 \end{align}
 whose derivation is detailed in Appendix~\ref{appendixB} for completeness.
 \subsection{Choosing Appropriate $\xi_n^{(1)}$ and $\xi_n^{(n)}$ to Simplify Bounds}
 Choose
 \begin{equation}
\log \xi_n^{(i)}\triangleq n\mathrm{C}\left(\tilde P^{(i,n)}\right) + \frac{\sqrt{n}\log n}{2\left(\tilde P^{(i,n)}+1\right)} \label{xiNchoice}
 \end{equation}
 for each $i\in\{1,2\}$. Combining~\eqref{converseProofEq7}, \eqref{converseProofEq9}, \eqref{converseProofEq10}, \eqref{converseProofEq11} and~\eqref{converseProofEq12} and recalling the definitions of~$\gamma_n^{(i)}$, $t_n$ and~$\xi_n^{(i)}$ in~\eqref{defGammaI}, \eqref{defTn} and~\eqref{xiNchoice} respectively, we have for each $i\in\{1,2\}$
 \begin{align}
  \Pr\left\{\sum_{k=1}^n i_k^{(i)}\ge \log \xi_n^{(i)}\right\} \le n^{-1}\left(1+2\tilde P^{(i,n)}n^{-1/2}\right)^{\frac{-n}{2}}e^{\sqrt{n}\tilde P^{(i,n)}+\frac{2\tilde P^{(i,n)}}{1+2\tilde P^{(i,n)}n^{-1/2}}} + 1-\varepsilon-n^{-1/2}.  \label{converseProofEq13}
 \end{align}
 Using the well-known inequality
 \begin{equation}
 \left(1+\frac{a}{m}\right)^{m} \le e^a \le  \left(1+\frac{a}{m}\right)^{m+a}   \label{converseProofEq14}
 \end{equation}
 for any $a>0$ and any $m>0$, which is shown in Appendix~\ref{appendixC} for completeness, we obtain for each $i\in\{1,2\}$
 \begin{align*}
 \left(1+2\tilde P^{(i,n)}n^{-1/2}\right)^{\frac{-n}{2}} &= \frac{\left(1+\frac{2\tilde P^{(i,n)}}{\sqrt{n}}\right)^{\sqrt{n}\tilde P^{(i,n)}}}{\left(1+\frac{2\tilde P^{(i,n)}}{\sqrt{n}}\right)^{\left(\frac{\sqrt{n}}{2}+\tilde P^{(i,n)}\right)\sqrt{n}}}\\
 & \le \frac{e^{2\left(\tilde P^{(i,n)}\right)^2}}{e^{\sqrt{n}\tilde P^{(i,n)}}},
 \end{align*}
 which implies from~\eqref{converseProofEq13} that
 \begin{align}
  \Pr\left\{\sum_{k=1}^n i_k^{(i)}\ge \log \xi_n^{(i)}\right\} \le  \frac{e^{2\tilde P^{(i,n)}\left(\tilde P^{(i,n)}+1\right)}}{n} + 1-\varepsilon-n^{-1/2}. \label{converseProofEq15}
 \end{align}
 By examining the definitions of $\tilde P^{(1,n)}$ and $\tilde P^{(2,n)}$ in~\eqref{defTildeP1n} and~\eqref{defTildeP2n} respectively and using~\eqref{defPinConverse*} and~\eqref{defRho*}, we conclude for each $i\in\{1,2\}$ that
 \begin{align*}
 \tilde P^{(i,n)} \le \left(\frac{2}{1-\varepsilon}\right)\left(\frac{P_1+P_2+\sqrt{P_1P_2}}{N_2}\right)
 \end{align*}
 and hence
 \begin{align}
 2\tilde P^{(i,n)}\left(\tilde P^{(i,n)}+1\right) \le \left(\frac{4}{1-\varepsilon}\right)\left(\frac{P_1+P_2+\sqrt{P_1P_2}}{N_2}\right)\left(\frac{2(P_1+P_2+\sqrt{P_1P_2})}{(1-\varepsilon)N_2}+1\right).  \label{converseProofEq16}
 \end{align}
 Combining~\eqref{converseProofEq15}, \eqref{converseProofEq16} and~\eqref{sufficientLarge2Converse}, we have for each $i\in\{1,2\}$
 \begin{align}
  \Pr\left\{\sum_{k=1}^n i_k^{(i)}\ge \log \xi_n^{(i)}\right\} \le 1-\varepsilon-\frac{1}{2\sqrt{n}}. \label{converseProofEq17}
 \end{align}
Combining~\eqref{converseProofEq2}, \eqref{converseProofEq4} and~\eqref{converseProofEq17} and recalling the definitions of~$i_k^{(1)}$, $i_k^{(2)}$, $\xi_n^{(1)}$ and $\xi_n^{(2)}$ in \eqref{defIk1}, \eqref{defIk2} and~\eqref{xiNchoice}, we have
\begin{align*}
\log M_n \le  n\mathrm{C}\left(\tilde P^{(i,n)}\right) + \frac{\sqrt{n}\log n}{2\left(\tilde P^{(i,n)}+1\right)} +\frac{1}{2}\log n + \log 2
\end{align*}
for each $i\in\{1,2\}$, which implies from the definitions of~$\tilde P^{(1,n)}$, $\tilde P^{(2,n)}$, $P_1^{(n)}$ and $P_2^{(n)}$ in~\eqref{defTildeP1n}, \eqref{defTildeP2n} and~\eqref{defPinConverse} that
\begin{align}
\log M_n \le  \frac{n}{2}\log\left(1+\frac{P_1+P_2+\rho\sqrt{P_1P_2}}{(1-\varepsilon-n^{-1/2})(N_2+N_3)}\right) + \sqrt{n}\log n +\frac{1}{2}\log n + \log 2 \label{converseProofEq18}
\end{align}
and
\begin{align}
\log M_n \le  \frac{n}{2}\log\left(1+\frac{(1-\rho^2)P_1}{(1-\varepsilon-n^{-1/2})N_2}\right) + \sqrt{n}\log n +\frac{1}{2}\log n + \log 2. \label{converseProofEq19}
\end{align}
Since
\begin{align*}
\frac{1}{1-\varepsilon-n^{-1/2}}& = \frac{1}{1-\varepsilon}+\frac{n^{-1/2}}{(1-\varepsilon)(1-\varepsilon-n^{-1/2})}\\
&\stackrel{\eqref{sufficientLarge1Converse}}{\le}\frac{1}{1-\varepsilon}+\frac{2n^{-1/2}}{(1-\varepsilon)^2},
\end{align*}
it follows from~\eqref{converseProofEq18} and~\eqref{converseProofEq19} and the inequality
\begin{equation*}
\log(1+a+b)\le \log(1+a)+b
 \end{equation*}
 for all $a,b>0$ based on Taylor's theorem that
\begin{align*}
\log M_n \le  \frac{n}{2}\log\left(1+\frac{P_1+P_2+\rho\sqrt{P_1P_2}}{(1-\varepsilon)(N_2+N_3)}\right) + \sqrt{n}\log n +\frac{\sqrt{n}}{(1-\varepsilon)^2}+\frac{1}{2}\log n + \log 2 
\end{align*}
and
\begin{align*}
\log M_n \le  \frac{n}{2}\log\left(1+\frac{(1-\rho^2)P_1}{(1-\varepsilon)N_2}\right) + \sqrt{n}\log n +\frac{\sqrt{n}}{(1-\varepsilon)^2}+\frac{1}{2}\log n + \log 2, 
\end{align*}
which implies from~\eqref{defRcutset} that
\begin{align}
\log M_n \le  R_{\text{cut-set}}\left(1-\rho^2, \frac{P_1}{1-\varepsilon}, \frac{P_2}{1-\varepsilon}\right)+\sqrt{n}\log n +\frac{\sqrt{n}}{(1-\varepsilon)^2}+\frac{1}{2}\log n + \log 2, \label{converseProofEq22}
\end{align}
which then implies from~\eqref{thmTempEq2} that
\begin{align}
R&\le  R_{\text{cut-set}}\left(1-\rho^2, \frac{P_1}{1-\varepsilon}, \frac{P_2}{1-\varepsilon}\right)\notag\\
&\stackrel{\eqref{defRho*}}{\le}  \max\limits_{0\le\alpha\le 1}R_{\text{cut-set}}\left(\alpha, \frac{P_1}{1-\varepsilon}, \frac{P_2}{1-\varepsilon}\right). \label{converseProofEq23}
\end{align}
Since~$\varepsilon\in(0,1)$ is arbitrary and $R$ is chosen to be an arbitrary $\varepsilon$-achievable rate, \eqref{convProofSt} follows from~\eqref{converseProofEq23}.

\appendices
\section{Detailed Calculations of~\eqref{lowerBoundonBProof} and~\eqref{lowerBoundonMProof}} \label{appendixA}
Consider the following facts due to~\eqref{defChannelInDefinition}, \eqref{defDistU}, \eqref{defDistV}, \eqref{defDistForDecodingBefore} and~\eqref{defDistForDecoding}:
\begin{align}
p_{Y_2|U,V}(y_2|u,v) &= \mathcal{N}\left(y_2  ; \sqrt{\tilde \alpha P_1^{(n)}} u + \sqrt{(1-\tilde \alpha) P_1^{(n)}}v \,,  N_2 \right), \label{defPY2GivenUV}\\
p_{Y_2|V}(y_2|v)& = \mathcal{N}\left(y_2 ; \sqrt{(1-\tilde \alpha) P_1^{(n)}}v \,,  (1-n^{-1/4}) \tilde \alpha P_1^{(n)}  + N_2 \right), \label{defPY2GivenV}\\
p_{Y_3|U,V}(y_3|u,v) &= \mathcal{N}\left(y_3 ; \sqrt{\tilde \alpha P_1^{(n)}} u + \left(\sqrt{(1-\tilde \alpha) P_1^{(n)}}+\sqrt{P_2^{(n)}}\right)v\,, N_2+N_3 \right), \\
p_{Y_3|V}(y_3|v)& =\mathcal{N}\left(y_3 ; \left(\sqrt{(1-\tilde \alpha)P_1^{(n)}} +\sqrt{P_2^{(n)}}\right) v\,, (1-n^{-1/4}) \tilde \alpha P_1^{(n)} +N_2+N_3 \right)
\end{align}
and
\begin{align}
p_{Y_3}(y_3) &=
\mathcal{N}\left(y_3;0,(1-n^{-1/4})\left(P_1^{(n)}+P_2^{(n)}+2\sqrt{(1-\tilde \alpha)P_1^{(n)}P_2^{(n)}}\right) + N_2+N_3 \right) \label{defPY3}
\end{align}
for all $(u,v,y_2,y_3)\in \mathbb{R}^4$. Recalling $\mathrm{C}(x)\stackrel{\eqref{defCapacityFunction}}{=}\frac{1}{2}\log(1+x)$ and defining $\mathrm{V}(x)\triangleq \frac{x}{1+x}$ for all $x\in \mathbb{R}_+$, we obtain from~\eqref{defPY2GivenUV}--\eqref{defPY3} that
 \begin{align}
\E\left[\log\left(\frac{p(Y_2|U,V)}{p(Y_2|V)}\right)\right]&=  \mathrm{C}\left(\frac{(1-n^{-1/4})\tilde \alpha P_1^{(n)}}{N_2}\right) , \label{eqnExpectaion2}\\
\Var\left[\log\left(\frac{p(Y_2|U,V)}{p(Y_2|V)}\right)\right]&=  \mathrm{V}\left(\frac{(1-n^{-1/4})\tilde \alpha P_1^{(n)}}{N_2}\right) \le 1,\label{eqnVar2}\\
\E\left[\log\left(\frac{p(Y_3|U,V)}{p(Y_3)}\right)\right]& = \mathrm{C}\left(\frac{(1-n^{-1/4})\left(P_1^{(n)}+P_2^{(n)}+2\sqrt{(1-\tilde \alpha)P_1^{(n)}P_2^{(n)}}\right)}{ N_2+N_3}\right),\label{eqnExpectaion3}\\
 \Var\left[\log\left(\frac{p(Y_3|U,V)}{p(Y_3)}\right)\right]& = \mathrm{V}\left(\frac{(1-n^{-1/4})\left(P_1^{(n)}+P_2^{(n)}+2\sqrt{(1-\tilde \alpha)P_1^{(n)}P_2^{(n)}}\right)}{ N_2+N_3}\right)\le 1, \label{eqnVar3}
 \\
  \E\left[\log\left(\frac{p(Y_3|V)}{p(Y_3)}\right)\right]& = \mathrm{C}\left(\frac{(1-n^{-1/4})\left( \sqrt{(1-\tilde \alpha)P_1^{(n)}} +\sqrt{P_2^{(n)}}\right)^2}{N_2+N_3 + (1-n^{-1/4})\tilde \alpha  P_1^{(n)}} \right), \label{eqnExpectation4}
 \\
&\hspace{-1.8 in}\text{and}\notag\\
\Var\left[\log\left(\frac{p(Y_3|V)}{p(Y_3)}\right)\right]& = \mathrm{V}\left(\frac{(1-n^{-1/4})\left( \sqrt{(1-\tilde \alpha)P_1^{(n)}} +\sqrt{P_2^{(n)}}\right)^2}{N_2+N_3 + (1-n^{-1/4})\tilde \alpha  P_1^{(n)}} \right)\le 1. \label{eqnVar4}
\end{align}
Consequently, \eqref{lowerBoundonBProof} follows from~\eqref{defB}, \eqref{eqnExpectation4} and~\eqref{eqnVar4}, and~\eqref{lowerBoundonMProof} follows from~\eqref{defM}, \eqref{eqnExpectaion2}, \eqref{eqnVar2}, \eqref{eqnExpectaion3}, \eqref{eqnVar3} and~\eqref{eqnVar4}.
\section{Detailed Derivation of~\eqref{converseProofEq12}}\label{appendixB}
Fix any $t>0$, any $P>0$, and any pair of random variables $(X^n, Z^n)$ such that $Z_k\sim\mathcal{N}(z_k;0,1)$ and $Z_k$ and $(X^{k}, Z^{k-1})$ are independent for each $k\in\{1, 2, \ldots, n\}$. We would like to prove that
\begin{equation}
\E\left[ e^{t\sum\limits_{k=1}^n\left(-PZ_k^2 + 2X_kZ_k+P\right)+\frac{2t^2}{1+2t P}\left(nP - \sum\limits_{k=1}^n X_k^2 \right)}\right] = (1+2t P)^{\frac{-n}{2}}e^{ntP + \frac{2nt^2P}{1+2t P}}, \label{st1AppendixB}
\end{equation}
which will then imply~\eqref{converseProofEq12} by relabelling the random variables. Consider the following chain of equalities for each $\ell\in\{1, 2, \ldots, n\}$:
\begin{align}
&\E\left[ e^{t\sum\limits_{k=1}^\ell\left(-PZ_k^2 + 2X_kZ_k\right)+\frac{2t^2}{1+2t P}\left(nP - \sum\limits_{k=1}^\ell X_k^2 \right)}\right] \notag\\
& = \E\left[\E\left[\left. e^{t\sum\limits_{k=1}^{\ell}\left(-PZ_k^2 + 2X_kZ_k\right)+\frac{2t^2}{1+2t P}\left(nP - \sum\limits_{k=1}^{\ell} X_k^2 \right)}  \right|X^{\ell-1}, Z^{\ell-1}\right]\right]\notag\\
& = \E\left[\left.\E\left[e^{t\sum\limits_{k=1}^{\ell-1}\left(-PZ_k^2 + 2X_kZ_k\right)+\frac{2t^2}{1+2t P}\left(nP - \sum\limits_{k=1}^{\ell-1} X_k^2 \right)}\E\left[\left. e^{t\left(-PZ_\ell^2 + 2X_\ell Z_\ell\right)-\frac{2t^2X_\ell^2}{1+2t P}}  \right|X^{\ell-1}, Z^{\ell-1}\right]\right|X^{\ell-1}, Z^{\ell-1}\right]\right]\notag\\
& \stackrel{\text{(a)}}{=} (2tP+1)^{\frac{-1}{2}} \E\left[ e^{t\sum\limits_{k=1}^{\ell-1}\left(-PZ_k^2 + 2X_kZ_k\right)+\frac{2t^2}{1+2t P}\left(nP - \sum\limits_{k=1}^{\ell-1} X_k^2 \right)}\right] \label{eqn1AppendixB}
\end{align}
where (a) follows from integrating $Z_\ell$ in the conditional expectation and using the facts that $Z_\ell\sim\mathcal{N}(z_\ell;0,1)$ and $Z_\ell$ and $(X^{\ell}, Z^{\ell-1})$ are independent. Applying~\eqref{eqn1AppendixB} recursively from $\ell=n$ to $\ell=1$, we obtain
\begin{equation*}
\E\left[ e^{t\sum\limits_{k=1}^n\left(-PZ_k^2 + 2X_kZ_k\right)+\frac{2t^2}{1+2t P}\left(nP - \sum\limits_{k=1}^n X_k^2 \right)}\right] = (1+2t P)^{\frac{-n}{2}}e^{\frac{2nt^2P}{1+2t P}},
\end{equation*}
which then implies~\eqref{st1AppendixB}.

\section{Derivation of~\eqref{converseProofEq14}}\label{appendixC}
Fix an $a>0$ and an $m>0$. Since
\begin{equation*}
\int_{1}^{1+\frac{a}{m}}\frac{1}{1+\frac{a}{m}} \,\mathrm{d} t \le \int_{1}^{1+\frac{a}{m}}\frac{1}{t} \,\mathrm{d} t \le \int_{1}^{1+\frac{a}{m}}1\, \mathrm{d} t,
\end{equation*}
it follows that
\begin{equation*}
\frac{a}{m+a} \le \log\left(1+\frac{a}{m}\right) \le \frac{a}{m},
\end{equation*}
which implies that
\begin{equation*}
e^\frac{a}{m+a} \le 1+\frac{a}{m} \le e^\frac{a}{m},
\end{equation*}
which then implies that
\begin{equation*}
\left(1+\frac{a}{m}\right)^{m} \le e^{a} \le  \left(1+\frac{a}{m}\right)^{m+a}.
\end{equation*}

\section*{Acknowledgment}
The authors would like to thank the Associate Editor Prof.\ Haim Permuter and the two anonymous reviewers for the useful comments that improve the presentation of this paper.


\ifCLASSOPTIONcaptionsoff
  \newpage
\fi

\end{document}